\documentclass[journal,comsoc]{IEEEtran}
\usepackage[T1]{fontenc}% optional T1 font encoding
\usepackage{times}
\usepackage{soul}
\usepackage{url}
\usepackage[hidelinks]{hyperref}
\usepackage[utf8]{inputenc}
\usepackage[small]{caption}
\usepackage{graphicx}
\usepackage{amsmath}
\usepackage{amsthm}
\usepackage{booktabs}
\usepackage{algorithm}
\usepackage{algorithmic}
\usepackage{multirow}
\usepackage{threeparttable}
\usepackage{amsfonts}
\usepackage{float}
\usepackage{url}
\usepackage{bm}
\usepackage{xcolor}
\usepackage{subfigure}
\usepackage{amssymb}

\newcommand{\cmmnt}[1]{}

\ifCLASSINFOpdf
\else
\fi
\usepackage{amsmath}
\usepackage[cmintegrals]{newtxmath}
\begin{document}
%\title{Controllable cross-speaker emotion transfer for end-to-end speech synthesis}
\title{Cross-speaker emotion disentangling and transfer for end-to-end speech synthesis}

% \author{Michael~Shell,~\IEEEmembership{Member,~IEEE,}
%     John~Doe,~\IEEEmembership{Fellow,~OSA,}
%     and~Jane~Doe,~\IEEEmembership{Life~Fellow,~IEEE}% <-this % stops a space
% \thanks{M. Shell was with the Department
% of Electrical and Computer Engineering, Georgia Institute of Technology, Atlanta,
% GA, 30332 USA e-mail: (see http://www.michaelshell.org/contact.html).}% <-this % stops a space
% \thanks{J. Doe and J. Doe are with Anonymous University.}% <-this % stops a space
% \thanks{Manuscript received April 19, 2005; revised August 26, 2015.}}

\author{Tao Li,
    Xinsheng~Wang,
    Qicong~Xie,
    Zhichao~Wang,
    Lei~Xie,~\IEEEmembership{Senior Member,~IEEE}
  % <-this % stops a space
 \thanks{Corresponding author: Lei Xie}
  
\thanks{Tao Li, Qicong  Xie, Zhichao Wang, and Lei Xie are with the School of Computer Science, Northwestern Polytechnical University, Xi’an 710072, China. Email: taoli@npu-aslp.org (Tao Li), xieqicong@mail.nwpu.edu.cn (Qicong Xie), zcwang\_aslp@mail.nwpu.edu.cn (Zhichao Wang),lxie@nwpu.edu.cn (Lei Xie)}

\thanks{Xinsheng Wang is with the School of Software Engineering, Xi’an Jiaotong University, Xi’an 710049, China, and also with the School of Computer Science, Northwestern Polytechnical University, Xi’an 710072, China. Email: wangxinsheng@stu.xjtu.edu.cn}
    }

%     ^------------^------------^----Do not want these spaces!


\markboth{Journal of \LaTeX\ Class Files,~Vol.~14, No.~8, August~2015}%
{Wang \MakeLowercase{\textit{et al.}}: Bare Demo of IEEEtran.cls for IEEE Communications Society Journals}
% *** name in the headers of peer review papers.       ***



\maketitle


\begin{abstract}
The cross-speaker emotion transfer task in text-to-speech (TTS) synthesis particularly aims to synthesize speech for a target speaker with the emotion transferred from reference speech recorded by another (source) speaker. During the emotion transfer process, the identity information of the source speaker could also affect the synthesized results, resulting in the issue of speaker leakage, i.e., synthetic speech may have the voice identity of the source speaker rather than the target speaker. This paper proposes a new method with the aim to synthesize controllable emotional expressive speech and meanwhile maintain the target speaker's identity in the cross-speaker emotion TTS task. The proposed method is a Tacotron2-based framework with emotion embedding as the conditioning variable to provide emotion information. Two emotion disentangling modules are contained in our method to 1) get speaker-irrelevant and emotion-discriminative embedding, and 2) explicitly constrain the emotion and speaker identity of synthetic speech to be that as expected. Moreover, we present an intuitive method to control the emotion strength in the synthetic speech for the target speaker. Specifically, the learned emotion embedding is adjusted with a flexible scalar value, which allows controlling the emotion strength conveyed by the embedding. Extensive experiments have been conducted on a Mandarin disjoint corpus, and the results demonstrate that the proposed method is able to synthesize reasonable emotional speech for the target speaker. Compared to the state-of-the-art reference embedding learned methods, our method gets the best performance on the cross-speaker emotion transfer task, indicating that our method achieves the new state-of-the-art performance on learning the speaker-irrelevant emotion embedding. Furthermore, the strength ranking test and pitch trajectories plots demonstrate that the proposed method can effectively control the emotion strength, leading to prosody-diverse synthetic speech.
\end{abstract}


% \fi
\IEEEpeerreviewmaketitle

\section{Introduction}
\label{sc:Introduction}
\IEEEPARstart{T}{ext}-to-speech (TTS) aims to generate human-like speech from text~\cite{Arik2017DeepVR,Sotelo2017Char2WavES,ping2017deep3,Yang2019ImprovingME}. In recent years, the development of attention-based sequence-to-sequence (seq2seq) neural models~\cite{Bahdanau2015NeuralMT,Sutskever2014SequenceTS}, brought a revolution to the TTS task, making it possible to synthesize natural speech via an end-to-end (E2E) way with $\textless text, audio \textgreater$ pairs as training data. Beyond synthesizing natural but prosaic speech, this paper aims to synthesize controllable and emotional expressive speech by transferring emotions from a source speaker, which is important for many voice-based human-computer interaction scenarios.

The early E2E speech synthesis models~\cite{wang2017tacotron,shen2018natural,Ren2019FastSpeechFR,yu2019durian,Ling2015DeepLF} focused on synthesizing prosaic speech that did not explicitly take the emotion into consideration. In contrast, the natural speech produced by our human beings is not only semantic but also expressive. As a part of the important information conveyed by human speech, emotional expressions are directly affected by the speaker's intentions that may lead to different emotions, e.g., $fear$, $angry$, $happy$, $sad$, $surprise$ and $disgust$. Therefore, how to present appropriate emotions in synthetic speech is important in building diverse audio generation systems and immersive human-computer interaction systems~\cite{Wang2018Style, Zhang2019LearningLR,Wang2017Uncovering,Bian2019Multi,xie2021multic}, and thus has been drawn much attention recently~\cite{Choi2019MultispeakerEA,Se2020Emotional,Liu2021ReinforcementLF,xie2021multi}. 

Based on whether the emotion can be transferred from another person, emotional speech synthesis can be roughly divided into \textit{same-speaker} and \textit{cross-speaker} scenarios. In the \textit{same-speaker} scenario, to synthesize the emotional speech of a single speaker, a straightforward way is to train a TTS model with categorized emotional data~\cite{Li2018EMPHASISAE,Zhu2019ControllingES} if sizable emotional data is available. Besides, there are also several other methods to achieve this goal, e.g., model adaptation on a base model using a small amount of emotional data~\cite{Ohtani2015Emotional,Inoue2017An} and code/embedding-based methods~\cite{Choi2019MultispeakerEA,Wu2018RapidSA,Li2018EmphaticSG}. However, the weakness of these \textit{same-speaker} methods is obvious. They can only be used to produce synthetic speech of the speaker that same to the one in the training data, which limits the generalization in real applications. In contrast, the \textit{cross-speaker} methods aim to transfer the emotion from a source speaker to the target speaker~\cite{Wang2018Style,Zhang2019LearningLR,Bian2019Multi,Skerry2018Towards,Xue2021CycleCN,wang2021enriching}, making the synthetic speech of the target speaker can express various emotions that do not exist exists in the database recorded by the target speaker.

A popular way to perform the \textit{cross-speaker} emotion transfer TTS is to learn speaker-irrelevant emotion representations, which can be extracted from reference audio with the desired emotion and then is used as a conditioning variable during the generating process~\cite{Zhang2019LearningLR,Se2020Emotional,Wu2019EndtoEndES,Jia2018TransferLF,Kwon2019AnES}. Reference Encoder~\cite{Skerry2018Towards}, global style tokens (GST)~\cite{Kwon2019EmotionalSS,Kwon2019AnES}, and variational autoencoder (VAE)~\cite{Zhang2019LearningLR,Kulkarni2020IMPROVINGLR} are commonly used strategies to extract the emotion representations. In addition to the learning of emotion embedding from reference speech, how to preserve the target speaker's voice in synthetic speech is also very important for the \textit{cross-speaker} emotion transfer TTS. To this end, some methods take the speaker embedding of the target speaker as extra conditioning information to the system~\cite{Skerry2018Towards,Gibiansky2017DeepV2}. However, as the transferred emotion is from speech uttered by another (source) speaker, identity information of this source speaker could also be transferred to synthesized speech, making synthetic speech sound somehow like uttered by the source speaker rather than the target speaker, i.e., the so-called \textit{speaker leakage} problem. To mitigate the speaker leakage, the existing approaches made a trade-off between the transfer quality and identity preservation, resulting in either transferred emotion in synthetic speech is not expressive enough, or synthetic speech is still suffering from the source speaker leakage~\cite{Karlapati2020CopyCatMF}.

Moreover, the incredibly challenging of recording an emotional expressive speech database with different labeled emotion categories and strengths makes it hard to provide a proper reference to deliver the desired emotion strength. Therefore, most existing \textit{cross-speaker} methods only focus on emotional speech synthesis without strength controlling. However, the same emotion category conveyed by our human beings' speech is appropriately presented by different `levels', e.g., \textit{very happy} and \textit{a little bit happy}, and how to control the emotion strength of synthesized speech is important for creating more realistic synthetic speech. While some emotion strength control methods have been proposed recently~\cite{Zhu2019ControllingES,Li2021ControllableET,Ferrari2008learning}, all of them are aimed at the \textit{same-speaker} scenario rather than the \textit{cross-speaker} scenario. In this work, the first effort to control the emotion strength in the \textit{cross-speaker} emotion transfer TTS task is conducted, in which, the strength control is realized by a flexible emotion scalar that bypasses the dependency on the database with manually-labeled emotion strengths.

To synthesize emotional expressive speech by transferring the emotion from reference audio and meanwhile maintain the target speaker identity in synthetic speech, a novel emotion transfer TTS approach is proposed in this paper. To be specific, the proposed method is a Tacotron2 \cite{shen2018natural} based TTS system with the emotion embedding as a conditioning variable to provide emotion information of reference speech. To obtain speaker-irrelevant and emotion-discriminative emotion embedding, an emotion disentangling module (EDM) is proposed. This proposed EDM consists of two encoders, i.e., emotion encoder and speaker encoder. During the training process, the emotion encoder is trained with the classification loss in terms of emotion categories, and meanwhile, the emotion embedding is constrained to be orthogonal to the speaker embedding produced by the speaker encoder with the same speech as input. This speaker encoder is trained with two loss functions: one is the classification loss in terms of speaker identities, and another one is gradient reversal layer (GRL)-based classification loss in terms of emotion categories, to make the learned speaker embedding be emotion-irrelevant and speaker-discriminative. Besides, to explicitly constrain the emotion category and speaker identity of synthetic speech, the proposed EDM is not only used for the learning of emotion embedding, but also used to calculate the corresponding loss with the synthesized Mel-spectrograms as input. An emotion matching loss calculated with a pair of emotion embeddings from reference audio and synthesized Mel-spectrograms respectively is used to ensure the consistency between the referenced emotion and the synthesized emotion. Moreover, as mentioned earlier that it is quite difficult to create an emotional expressive speech database with different emotion categories and strengths, therefore, in this paper we also aim to answer the question that whether it is possible to control the emotion strength of synthetic speech that synthesized by a model trained with emotional speech database without manually-annotated strength information. To this end, a scalar value-based method is introduced to adjust the emotion strength carried by the learned emotion embedding. 

Preliminary work of this paper was presented in~\cite{Li2021ControllableET}, in which the model was designed for the \textit{same-speaker} emotion transfer TTS, and the training of the emotion encoder and the speaker encoder in that model were not constrained with the orthogonality relation between emotion embeddings and speaker embeddings. In this paper, we improved the model in~\cite{Li2021ControllableET} with the proposed EDM and also extended it to the task of \textit{cross-speaker} emotion transfer TTS. To sum up, the main contributions of this work are as follows:
\begin{itemize}
    \item We propose a novel model for the task of \textit{cross-speaker} emotional transfer TTS. To solve the speaker leakage problem, an emotion disentangling module (EDM) is proposed in this paper to learn speaker-irrelevant emotion embeddings and emotion-irrelevant speaker embeddings at the same time with the orthogonal constraint.
    \item For the first time, the effort to control the emotion strength in the \textit{cross-speaker} emotion transfer TTS is conducted, and a simple emotion strength control method is proposed.
    \item Extensive experiments show that the proposed method outperforms other state-of-the-art methods, i.e., GST-based and VAE-based methods, on the \textit{cross-speaker} emotion transfer TTS, and also demonstrate that the proposed method is able to properly control the emotion strength for synthetic speech.
\end{itemize}

The rest of this paper is organized as follows. Section~\ref{sc:related work} reviews related work.  Section~\ref{sc:method} introduces the proposed method. Section \ref{sc:experiments} describes the experiments and setup in detail. Section \ref{sc:results} presents the experimental results. Section \ref{sc:discussion} discusses the performance and limitation of the proposed method, and also the possible future research direction. Finally, the paper concludes in Section \ref{sc:conclusion}. Examples of synthesized speech can be found on the project page\footnote{The synthesized utterances can be found from https://silyfox.github.io/multispkemotion/ \label{ft:homepage}}.

\section{Related work}
\label{sc:related work}
Emotion transfer in TTS shares similar methods with other kinds of style transfer, e.g., prosody transfer in TTS, as emotions are expressed in prosodic aspects of speech. Here, for clarity, all of them are referred to as style transfer. In this section, works related to style transfer and also controllable cross-speaker emotion transfer are reviewed.

\subsection{Style transfer TTS}

Given reference audio, style transfer TTS is to synthesize speech with the style learned from reference audio. Inspired by the good performance of the Tacotron series models~\cite{wang2017tacotron,shen2018natural} on synthesizing natural speech, many efforts have been conducted to extend the Tacotron structures to style transfer TTS~\cite{Skerry2018Towards,Wu2018FeatureBA,Bian2019Multi}.

The early method to achieve style transfer is to integrate the Tacotron model with an extra trainable audio encoder, referred to as Reference Encoder, which encodes the reference audio as a fixed-length vector that works as conditioning information of the expected style~\cite{Skerry2018Towards}. The following works~\cite{Wang2018Style,Se2020Emotional} extend the Reference Encoder module by adding a style token layer, which is composed of an embedded library called Global Style Tokens (GST). GST learns a latent high-dimensional representation for style transfer that implicitly contains style information. Besides, the Variational Auto-Encoder module~\cite{Kingma2014AutoEncodingVB} also has been shown good performance on learning the potential representation of the style in the style transfer TTS task ~\cite{Zhang2019LearningLR}. Recently, some variants based on the mentioned Reference Encoder, GST, or VAE are proposed to further improve the performance on the style transfer TTS task~\cite{Bian2019Multi,2019Multi,Ma2019NeuralTS,Kulkarni2020TransferLO,Valle2020MellotronME,Sorin2020PrincipalSC,Li2021TowardsMS}. For instance, in~\cite{Sorin2020PrincipalSC}, a style-related latent space is learned in an unsupervised way by Reference Encoder, and is then transformed to a low-dimensional space with Principal Component Analysis (PCA) to disentangled style from text and speaker-related information. In~\cite{Li2021TowardsMS}, a multi-scale Reference Encoder is employed to extract the global-scale utterance-level and the local-scale quasi-phoneme-level style features of the target speech, with the goal to improve the expressiveness of the synthesized speech.

However, most style modeling methods aggregate all style-related aspects, e.g., pitch, duration, emotion and accent, into one hidden style representation, making the learned system can only transfer the \textit{average} expressiveness. To control different style aspects independently, Bian et al.~\cite{Bian2019Multi} proposed a multi-reference TTS stylization strategy based on GST-Tacotron~\cite{Wang2018Style} and an intercross training scheme, in which different style dimensions, such as emotion and speaker, are disentangled and transferred independently. Subsequently, Whitehill et al.~\cite{2019Multi} improved the performance of the multi-reference model on disjoint-datasets by unpaired training strategy and adversarial cycle consistency scheme.

While numerous methods have been proposed to improve the style transfer performance as mentioned above, the lack of explicit constraints on the final synthesized speech makes them suffer from the conflict between the well-transferred styles and identity preservation of the target speaker in the cross-speaker scenario~\cite{Karlapati2020CopyCatMF}. Furthermore, how to create flexible stylized speech with controllable transferred strength was not considered.

\subsection{Emotion strength control}
The emotion strength control aims to synthesize emotional speech of different strengths as expected. In the same-speaker scenario, an unsupervised ranking function is learned from the emotional dataset based on relative attributes scheme~\cite{Ferrari2008learning}, and each training sample is assigned a relative emotion strength~\cite{Zhu2019ControllingES}, which is then used as a strength label to condition the Tacotron model. At inference time, emotion expressions can be easily controlled by a discrete one-hot vector presenting emotion category and a continuous simple scalar indicating emotion strength. In~\cite{Um2020EmotionalSS}, an interpolation technique was proposed to control the strength of the target emotion that can be gradually changed from weakness to neutral. Some work uses TD-PSOLA~\cite{Akanksh2015InterconversionOE,Kannan2021VoiceCU} kind of techniques to manually modify prosodic components, such as pitch contour and duration, resulting in different perceived emotions and could also modify the emotion strength. 

Recently, we proposed a controllable emotion speech synthesis approach~\cite{Li2021ControllableET} to deliver the emotion accurately and control the emotion strength flexibly. Specifically, we modified the Reference Encoder structure with two emotion classifiers to enhance the emotion-discriminative ability of the emotion embedding and the predicted speech representation, i.e., Mel-spectrum. Besides, an emotion matching loss~\cite{Johnson2016Perceptual,GatysA} was adopted to minimize the difference between the generated and reference Mel-spectrum in terms of the emotion. During the inference, the strength of the synthetic speech can be easily controlled by adjusting a scalar to the emotion embedding. However, we found that this approach can not avoid the source speaker leakage in the cross-speaker scenario. In this paper, we use our preliminary method~\cite{Li2021ControllableET} as the backbone and further improve it for the cross-speaker scenario. 

\begin{figure*}[htb]	
	\centering
	\includegraphics[width=18cm]{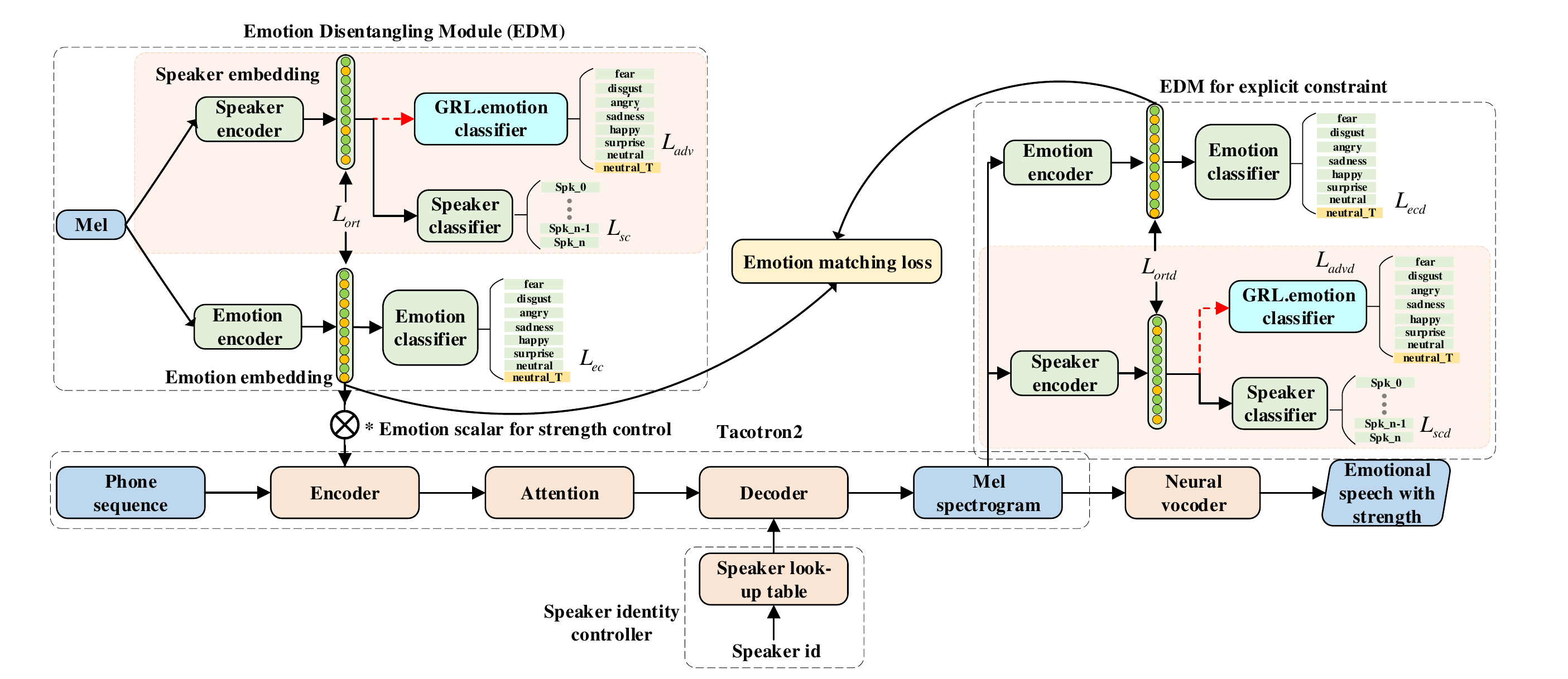}
    %\captionsetup{belowskip=-18pt}
	\caption{The architecture of the proposed cross-speaker model. The input text is represented as the phone sequence, and speech is represented by Mel-spectrogram which can be converted to waveform signal via a vocoder. The input ``Mel" is from reference audio to provide the speaker-irrelevant emotion embedding via the emotion encoder. Two EDMs are just used during the training processing, and only the emotion encoder with ``Mel" as input is kept during the inference.}
	\label{fig:frame1}
\end{figure*}

\section{Methodology}
\label{sc:method}
The illustration of the proposed architecture for the controllable cross-speaker emotion transfer is shown in Figure~\ref{fig:frame1}. We utilize a modified Tacotron2~\cite{shen2018natural} as the backbone of this model. Following the Reference Encoder-based methods \cite{Skerry2018Towards,Sorin2020PrincipalSC}, we also take a Reference Encoder, named as emotion encoder, to get the emotion embedding that conveys emotion information from reference speech to the TTS system. As mentioned earlier, speaker information preserved in this emotion embedding could lead to speaker leakage. Therefore, learning speaker identity irrelevant emotion embedding is very important for the \textit{cross-speaker} emotion transfer TTS task. To this end, our emotion encoder is trained with a proposed Emotion Disentangling Module (EDM, see the upper-left region within the dashed box), which is able to disentangle speaker information from the emotion embedding. This EDM is also used for the emotion encoder with synthesized Mel-spectrum as input to constrain the emotion and identity of synthesized speech (see the upper-right region within the dashed box). The target speaker's identity is provided by the Identity Controller. In this section, the Tacotron2-based backbone, EDM, and the Identity Controller will be introduced respectively. Besides, a flexible emotion scalar is used to control the emotion strength, which will also be introduced in this section. 

Note that all the proposed modules and the TTS backbone, are trained together. The objective functions of each module will be also introduced in this section.

\subsection{Tacotron2-based backbone}

The backbone of the proposed method is based on Tacotron2~\cite{shen2018natural}, which is a state-of-the-art attention-guided seq2seq TTS model that consists of the encoder, decoder, and attention mechanism module. With the input of a phone sequence, the encoder produces a sequence of intermediate representations that work as input via the attention module to the decoder to produce the speech representations, i.e., Mel-spectrogram. The final audio signal is obtained by a neural vocoder with the predicted spectrograms as input. 

Different from the vanilla Tacotron2, here, we replace the encoder in Tacotron2 with that in Tacotron, which consists of a pre-net and a CBHG module~\cite{Lee2017Fully}.  Besides, instead of the location-sensitive attention mechanism used in Tacotron2, the GMM attention mechanism~\cite{Battenberg2020LocationRelativeAM} is adopted in this work. The Tacotron2 is optimized to minimize the mean absolute error (MSE) of predicted Mel-spectrograms and ground-truth Mel-spectrograms. The loss function that to train the Tacotron2 is referred to as ${\cal L}_{taco}$.

\subsection{Emotion Disentangling Module (EDM)}
By disentangling speaker information from the emotion embedding, the EDM is to learn the emotion encoder that can obtain the identity-irrelevant emotion embedding. The EDM consists of two encoders, i.e., speaker encoder and emotion encoder. The emotion embedding $e_i$ produced by the emotion encoder should be discriminative for emotion categorization and be unrelated to the speaker embedding $s_i$ extracted from the same audio. In this section, we will introduce the architecture of the proposed emotion encoder and the strategies to train this encoder. Besides, the details of the speaker encoder and also the way to learn the emotion-irrelevant speaker embedding $s_i$ will also be introduced.

\subsubsection{Emotion encoder}
\label{sc:emotion encoder}
Similar to the emotion encoder architecture in~\cite{Skerry2018Towards}, here, the emotion encoder consists of six 2D convolutional layers, a GRU layer, three fully connected (FC) layers. Only the last GRU state of the GRU layer is taken as the global feature that works as the input of the FC layers. The final emotion embedding $e_i$ is represented by a 256-dimensional vector.

To make the emotion embedding $e_i$ discriminative on distinguishing different emotions, a classification loss in terms of emotion categories is adopted. Specifically, the emotion embedding $e_i$ is fed into an emotion classifier, which consists of an FC layer and a softmax layer. Instead of transferring the emotion embedding from 256 dimensions to the dimension that same to the number of source emotion categories, i.e. 7 in this paper, one more dimension is outputted from the FC layer, which presents the style of the target speaker. In this way, the style of the target speaker is treated as an irrelevant emotion category, which is inspired by the fact that even with the same emotion category, slight emotion differences would exist between speech from different persons. For simplicity, here, we denote this neutral emotion from the target speaker as \textit{neutral\_T}. The softmax layer is to produce the probability of eight emotion types, i.e., $neutral$, $happy$, $surprise$, $angry$, $disgust$, $fear$, $sad$, and $ neutral\_T$. The corresponding objective function is then defined as the negative log probability of $P( {{y_i}\mid {e_i}})$
\begin{equation}
    {{\cal L}_{ec}} =  - \sum\limits_{i = 1}^n {\log } P\left( {{y_i}\mid {e_i}} \right)
\end{equation}
where $n$ is the batch size, $y_i$ is the emotion label of the emotion embedding $e_i$, and $P( {{y_i}\mid {e_i}})$ is possibility of $e_i$ belonging to the label $y_i$.

To make the emotion embedding unrelated to the speaker information, an orthogonality loss is proposed to make the emotion embedding orthogonal to the speaker embedding. To be specific, assume that we have an emotion-irrelevant speaker embedding $s_i$ that is extracted from the same audio of $e_i$, the goal is to minimize the following orthogonality loss, which is defined as
\begin{equation}
    {{\cal L}_{ort}} = \sum\limits_{i = 1}^n {\left\| {s_i^T{e_i}} \right\|_F^2} 
\label{eq:orth_loss}
\end{equation}
where $\left \| \cdot \right \|_{F}$ is the Frobenius norm. The learning of the emotion-irrelevant speaker embedding $s_i$ will be introduced in the following subsection.

\subsubsection{Learning of emotion-irrelevant speaker embedding}
\label{sc:Learning of emotion-irrelevant speaker embedding}
A speaker encoder that has the same architecture as the emotion encoder is used to obtain the speaker embedding. The speaker embedding produced by this speaker encoder should be 1) discriminative on distinguishing identities, and 2) with no information related to the emotion. To this end, two loss functions are used to optimize the speaker encoder. One is the classification loss to make the obtained embedding speaker-discriminative. Another one is an adversarial loss to make the obtained embedding emotion-irrelevant.

\textbf{Classification loss} is similar to the classification loss of the emotion encoder. Specifically, a speaker classifier that consists of an FC layer and a softmax layer is used to get the probability of a speaker embedding $s_i$ belonging to the corresponding identity label $l_i$, and the corresponding loss is defined as 
\begin{equation}
    {{\cal L}_{sc}} =  - \sum\limits_{i = 1}^n {\log } P\left( {{l_i}\mid {s_i}} \right)
\end{equation}
where $P( {{l_i}\mid {s_i}})$ is possibility of $s_i$ belonging to the label $l_i$.

\textbf{Adversarial loss} is to make the speaker embedding $s_i$ be emotion-irrelevant. Instead of training the model in an alternative way to make the speaker encoder cannot produce emotion-discriminative embeddings, here, a gradient reversal layer (GRL) is adopted between the speaker encoder and an emotion-based classifier. So that we can minimize the classification loss of this emotion-based classifier to reversely optimize the speaker encoder on the emotion classification task. Therefore, the loss function is defined as
\begin{equation}
    {{\cal L}_{adv}} =  - \sum\limits_{i = 1}^n {\log } P\left( {{y_i}\mid {s_i}} \right)
\end{equation}
where $P( {{y_i}\mid {s_i}})$ is the possibility of the speaker embedding $s_i$ extracted from speech with the emotion category of $y_i$.

%\subsubsection{Objective function}
\subsubsection{EDM objective function}
Regarding the emotion disentangling module, the total objective function is defined as
\begin{equation}
{{\cal L}_{edm}} = {{\cal L}_{ec}} + \alpha {{\cal L}_{ort}} + \beta {{\cal L}_{sc}} +  (1-\beta){{\cal L}_{adv}}
\end{equation}
where $\alpha$ and $\beta$ are two hyper-parameters to balance the weights of different losses. Empirically, based on the experimental comparisons, $\alpha$ and $\beta$ are set as $0.02$ and 0.5 respectively.

\subsubsection{EDM for explicit constraint}
In addition to obtaining the emotion embeddings that concatenated with the encoder's outputs of Tacotron2-based backbone, the EDM is also used for the backbone's decoder (the upper-right dashed box in Fig. \ref{fig:frame1}) to ensure 1) source emotion and speaker identity is presented by the synthetic speech, and 2) emotion consistency between reference and synthetic speech (see section~\ref{sc:emtloss}). With the EDM, we can explicitly restrain synthesized speech as we expected. Note that parameters of two EDM (one for the encoder and another for the decoder) are not shared, while they have the 
similar objective function. We denote the objective function of the EDM for the decoder as ${{\cal L}_{edmd}}$.

\subsubsection{Emotion matching loss}
\label{sc:emtloss}
Besides supervising the emotion encoder in a high-level way, i.e., with the emotion classifier, we argue that it is also important to ensure the emotion representation of synthetic speech be similar to that of reference speech. To this end, an emotion matching loss is introduced to constrain the emotion embedding of synthesized Mel-spectrograms to be close to the emotion embedding of reference speech. Details of this emotion matching loss can be found in our previous work \cite{Li2021ControllableET}, and here we refer to this loss as ${{\cal L}_{emo}}$.

\subsection{Emotion strength control}

The emotion embedding in our proposed model is a collection of CNN output sequences, and the essence is the extraction and quantification of features. Therefore, each value can be considered as the strength of emotion-related features, and the strength of transferred emotion to the target speaker can be easily controlled by adjusting its value without being affected by the source speaker. In this paper, we use an emotion scalar to multiply the emotion embedding to control emotion transfer strength at the inference stage. This is similar to the degree control in image style transfer~\cite{Johnson2016Perceptual}. Note that this scalar always equals $1$ during the training process.

\subsection{Speaker identity controller}

While the emotion-irrelevant speaker representation can be obtained by the speaker encoder of the EDM module, the learned speaker representation is detrimental and unstable to control the speaker's identity due to the adversarial training strategy (see Section \ref{sc:Learning of emotion-irrelevant speaker embedding}). Therefore, we adopt a trainable speaker embedding table~\cite{Skerry2018Towards,Xue2021CycleCN} (see the speaker identity controller module in Fig.~\ref{fig:frame1}) to extract the speaker embedding of the target speaker. In detail, with the target speaker ID, a 128-dimensional speaker embedding can be produced by a speaker look-up table. Then this speaker embedding is concatenated with the decoder's input in the frame-level to provide the target speaker's information. Note that, in this paper, all mentioned speaker embeddings except for that in this subsection refer to representations extracted by the speaker encoder in EDM.

\subsection{Final objective function}

All modules introduced in the previous sections are trained together. The final objective function of the proposed model is defined as:
\begin{equation}
    {\cal L} = {{\cal L}_{taco}} + {{\cal L}_{edm}} + {{\cal L}_{edmd}} + {{\cal L}_{emo}}
\end{equation}

\begin{table*}[h]
\centering
\caption{Disjoint database for the cross-speaker emotion transfer TTS.}
\setlength{\tabcolsep}{3.5mm}
\label{tab:database}
\begin{tabular}{ccccccccccc}
\toprule
\multirow{2}{*}{Database} & \multirow{2}{*}{Speaker} & \multirow{2}{*}{Language} & \multicolumn{7}{c}{Emotion}   & \multirow{2}{*}{Role} \\
 &    &    & neutral & happy & angry & disgust & fear & surprise & sadness &   \\ \midrule
DB\_1   & Adult female  & Mandarin    & \checkmark   &    &     &    &   &   &    & Target speaker   \\
AIC   & Imitated girl  & Mandarin  & \checkmark  & \checkmark    & \checkmark    & \checkmark   & \checkmark   & \checkmark  & \checkmark    & Source speaker    \\ 
DB\_6   & Adult female  & Mandarin  &   & \checkmark    & \checkmark    & \checkmark   & \checkmark   & \checkmark  & \checkmark    & Source speaker    \\ \bottomrule
\end{tabular}
\end{table*}

%\vspace{-5pt}
\section{Experiments}
\label{sc:experiments}
To evaluate the performance of the proposed method on the controllable emotion transfer task, three Mandarin corpora are adopted (see Section \ref{sc:database}) for extensive experiments. The synthesized results are evaluated via the human rating experiment subjectively and also analyzed objectively. In this section, the databases adopted in the experiment, experimental setups, and human rating method will be introduced.

\subsection{Database}
\label{sc:database}

With the goal to transfer emotion styles from a source speaker to a disjoint target speaker, a database that consists of at least two speakers is required, in which one is the source speaker to provide the emotional speech and another provides the target speaker who does not have the emotional speech. Table~\ref{tab:database} shows the databases that we used to evaluate the performance of our proposed method on the \textit{cross-speaker} emotion transfer TTS task. Note that, here, two emotional databases, i.e., AIC and DB\_6, are used for the convenience of evaluating the ability of the proposed method on learning speaker-irrelevant emotion embeddings (see Section~\ref{sc:embedding visualization}). Considering the significant difference voices of speakers in DB\_1 and AIC, AIC is taken as the source database in the subjective evaluation, which makes it easy for participants to distinguish the voice difference, and also can demonstrate the ability of the proposed method on transferring emotion between two quite different voices. The details of these databases are as follows. 

\textbf{DB\_1} is a publicly available adult female neutral corpus\footnote{The dataset is available at \url{http://www.data-baker.com/hc_znv_1.html}}, which contains about 12 hours of speech utterances recorded in a professional studio. The training set and the test set are composed of $9900$ and $100$ utterances respectively. In our experiments, the speaker from this database works as the target speaker.
    
\textbf{Adults-imitated children (AIC)} is a high-quality emotional speech corpus containing 14-hour of recordings by a professional Chinese actress~\cite{Li2021ControllableET}. She imitates a little girl to perform seven categories of emotion (\textit{neutral}, \textit{happy}, \textit{angry}, \textit{disgust}, \textit{fear}, \textit{surprise}, and \textit{sadness}), which consists of 6000 neutral utterances and 620 utterances for each emotion type. In this paper, the speaker of this database works as the source speaker to provide reference speech with different emotions during inference. 

\textbf{DB\_6} is a female emotional corpus with around 14 hours' recordings. It contains all the emotion styles as that in \textit{AIC} except for neutral speech. There are 2000 speech utterances for each emotion category.

\subsection{Human perceptual rating experiment} 
The goal of the emotion transfer TTS is to synthesize speech with the emotion of reference speech and the voice of the target speaker, making the evaluation a subjective task. Therefore, a human perceptual rating experiment is performed to evaluate the synthetic speech in terms of emotion similarity (between synthetic speech and reference speech) and speaker similarity (between synthetic speech and target speaker's voice). Following the typical differential mean opinion score (DMOS) test method, participants are asked to rate given speech a score ranging from one to five for its emotion similarity or speaker similarity. The rating criteria is: \textit{bad = 1}; \textit{poor = 2}; \textit{fair = 3}; \textit{good = 4}; \textit{great = 5}, in 0.5 point increments.

We randomly select ten sentences from the test set of the target speaker's database to synthesize speech with six kinds of emotions respectively, resulting in 60 testing utterances. Twenty participants who are native Chinese took part in this experiment, and the final score for each utterance is the average of scores rated by all participants for this sample. In all tests, the results are associated with 95$\%$ confidence intervals. In addition to the DMOS evaluation, to evaluate the strength of synthetic speech, a ranking-based subjective evaluation was performed. The details can be found in Section \ref{sc:emtion_control}.

\begin{table*}[t]
 \caption{Comparison of our proposed method with Mspk-GST and Mspk-VAE in terms of speaker and emotion similarity DMOS with confidence intervals of 95$\%$. The higher value means better performance, and the bold indicates the best performance out of three models in terms of each emotion.}
 \label{tab:transfer}
\setlength{\tabcolsep}{6mm}
 \centering
\begin{tabular}{l|lll|lll}
\toprule
\multicolumn{1}{c|}{\multirow{2}{*}{Emotion}} & \multicolumn{3}{c|}{Speaker similairty DMOS}                                                  & \multicolumn{3}{c}{Emotion similarity DMOS}                                                \\ \cmidrule{2-7} 
\multicolumn{1}{c|}{}                         & \multicolumn{1}{c}{Mspk-GST} & \multicolumn{1}{c}{Mspk-VAE} & \multicolumn{1}{c|}{Proposed} & \multicolumn{1}{c}{Mspk-GST} & \multicolumn{1}{c}{Mspk-VAE} & \multicolumn{1}{c}{Proposed} \\ \midrule
fear     & \multicolumn{1}{l}{\bf{4.0}$\pm$\bf{0.058}}   & \multicolumn{1}{l}{3.74$\pm$0.079}  & \multicolumn{1}{l|}{3.95$\pm$0.062}  & \multicolumn{1}{l}{2.41$\pm$0.079}         & \multicolumn{1}{l}{3.36$\pm$0.087}         & \multicolumn{1}{l}{\bf{3.75}$\pm$\bf{0.073}}         \\
disgust    &\bf{4.09}$\pm$\bf{0.052} &3.73$\pm$0.061 &4.01$\pm$0.055 &2.69$\pm$0.068 &2.93$\pm$0.081 &\bf{3.55}$\pm$\bf{0.075} \\ 
angry      &\bf{3.95}$\pm$\bf{0.054} &3.60$\pm$0.058 &3.90$\pm$0.056 &2.73$\pm$0.069 &\bf{3.68}$\pm$\bf{0.074} &3.62$\pm$0.07 \\
sadness    &\bf{3.88}$\pm$\bf{0.059} &3.27$\pm$0.081 &3.79$\pm$0.064 &2.8$\pm$0.069 &3.85$\pm$0.065 &\bf{3.94}$\pm$\bf{0.055} \\
happy      &\bf{3.86}$\pm$\bf{0.053} &3.13$\pm$0.077 &3.76$\pm$0.052 &2.92$\pm$0.055 &\bf{3.72}$\pm$\bf{0.064} &3.70$\pm$0.06 \\
surprise   &\bf{4.01}$\pm$\bf{0.05} &3.54$\pm$0.062 &3.90$\pm$0.055 &2.5$\pm$0.071 &2.85$\pm$0.062 &\bf{3.69}$\pm$\bf{0.068} \\ \midrule
average    &\bf{3.97}$\pm$\bf{0.054} &3.50$\pm$0.069 &3.89$\pm$0.057 &2.68$\pm$0.068 &3.40$\pm$0.072 &\bf{3.71}$\pm$\bf{0.066} \\ \bottomrule
\end{tabular}
\end{table*}

\subsection{Experimental setups}
\label{exset}
%\vspace{-2pt}

There are a total of 31,720 speech utterances in the database, each of which has a duration between 1 and 9 seconds. All the speech utterances are down-sampled to 16 kHz, and are then represented by Mel-spectrograms extracted with $50$ ms frame length and $12.5$ ms frameshift. A grapheme-to-phoneme (G2P) module is used to convert the input sentences into phone sequences, which then work as the input to the proposed model, resulting in a predicted Mel-spectrogram. During the inference stage, the multi-band WaveRNN~\cite{yu2019durian} is adopted as the neural vocoder to reconstruct waveform from the predicted Mel-spectrograms.

For the emotion strength control, although the emotion scalar can be set continuously to control the emotion strength for each emotion category, it is hard for listeners to distinguish the subtle emotion differences. Therefore, we set three levels, i.e., weak, medium, and strong, for the evaluation. Different from the \textit{same-speaker} scenario~\cite{Li2021ControllableET}, considering the fact that the emotional information conveyed by the emotion embedding tends to be weakened in the process of squeezing out the source speaker's timbre information, which could make the emotional expressiveness become insipid in synthetic speech~\cite{Karlapati2020CopyCatMF}, we set the scalar as $1$ to represent the weak emotion strength and use $2$ and $3$ to represent relatively medium and strong strength. While the larger scalar means the stronger emotion strength, the scalar cannot be infinite. We found that a larger strength scalar does not always bring stronger emotion transfer with high quality. It is easy to be explained: even for real emotional speech, much strong emotional expressive speech may bring changes to the speaker's voice, which could affect the listener's judgment on the identity of the speaker. For the comparison with other methods in Section \ref{sc:comparsion with other methods}, the emotion scalar is set as $1$. Since the transfer of the neutral emotion will not change the style of the target speaker, the transfer of the neutral emotion will not be conducted in the follow-up experiment.

\subsection{Compared methods}
As mentioned in the related work in Section~\ref{sc:related work}, GST~\cite{Wang2018Style} and VAE~\cite{Zhang2019LearningLR} are two state-of-the-art strategies that are used in the speech style transfer task. Here, to show the superiority of our proposed method, these two strategies are also adopted to compare with our method. For fairness, we replace the Tactoron with the modified Tacotron2 in the GST-based and VAE-based TTS model, and the same speaker controller module in our proposed method is used in these two GST-based and VAE-based models to provide target speaker information. The two models are referred to as Mspk-GST and Mspk-VAE respectively hereafter for simplicity.

\section{Experimental results}
\label{sc:results}
In this section, experimental results for comparison with other methods, emotion strength control in the cross-speaker emotion transfer task, and ablation study are presented. The corresponding demos can be found on the project page\textsuperscript{\ref{ft:homepage}}, and we recommend readers listen to those demos.

\subsection{Comparison with other methods}
\label{sc:comparsion with other methods}
\subsubsection{Performance on the emotion transfer TTS}

To compare the proposed method with Mspk-GST and Mspk-VAE methods, performances of these three models on the cross-speaker emotion transfer are evaluated in terms of the speaker similarity and emotion similarity via the human rating experiments. The results are shown in Table~\ref{tab:transfer}.  

As can be seen in Table~\ref{tab:transfer}, regarding the speaker similarity, the GST-based method Mspk-GST achieves the best performance on preserving the target identity in the synthesized speech in terms of transferring all emotions. However, the emotion expressiveness performance of Mspk-GST is much worse than both Mspk-VAE and the proposed method. Specifically, in terms of the emotion similarity DMOS averaging on all emotion styles, the score of Mspk-VAE and our proposed method are 27\% and 38.4\% relatively higher than that achieved by Mspk-GST, respectively. The bad performance of Mspk-GST on the emotion expressiveness could be caused by the inadequate emotion representation produced by the GST module. This reason also brings a weak impact on the speaker identity of synthesized speech, which leads to higher performance on preserving the target speaker identity. This phenomenon can be easily explained by an extreme case: when the emotion embedding contains no information of reference audio, the target speaker identity in synthesized speech will not be affected. While Mspk-GST achieves the best performance on the speaker preservation, the terrible scores on emotion similarity DMOS indicate the weakness of Mspk-GST on the \textit{cross-speaker} emotion transfer TTS.

Compared with Mspk-GST, Mspk-VAE achieves better performance on the emotion expressiveness in terms of all emotion categories. However, the improvement is not always significant for all emotion categories. Specifically, when the emotion is reflected by the speaking speed and stress (such as $surprise$, $disgust$ and $fear$), the emotion expressiveness performance is still unsatisfactory. Meanwhile, when emotion categories are likely reflected by the changes of the source speaker's timbre (such as $sadness$, $happy$ and $angry$), Mspk-VAE gets an obvious drop in the performance of speaker preservation. For instance, in terms of the emotion \textit{happy}, the speaker similarity DMOS is 18.9\% and 16.8\% relatively lower than that achieved by Mspk-GST and the proposed method respectively. These results demonstrate that the VAE-based method Mspk-VAE is hard to achieve a balance between the target speaker preservation and source emotion transfer. 

\begin{figure*}[ht]
	\centering
	\begin{minipage}{\linewidth}
		
		\begin{minipage}{0.32\linewidth}
			\centerline{\includegraphics[width=\textwidth]{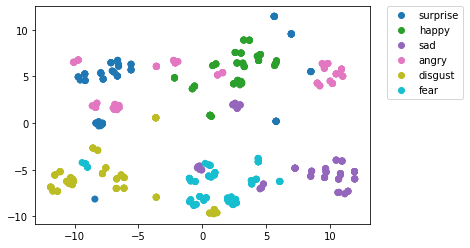}}
			\centerline{(a)}
		\end{minipage}
		\hfill
		\begin{minipage}{0.32\linewidth}
			\centerline{\includegraphics[width=\textwidth]{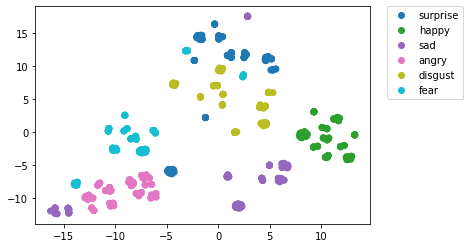}}
			\centerline{(b)}
		\end{minipage}
		\hfill
		\begin{minipage}{0.32\linewidth}
			\centerline{\includegraphics[width=\textwidth]{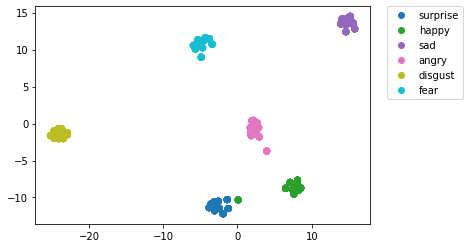}}
			\centerline{(c)}
		\end{minipage}
		\vfill
		%\centering{(a) Surprise}
	\end{minipage}
	\caption{Emotion distribution of the emotion embeddings created different models (a) GST, (b) VAE, and (c) Proposed EDM module. For ease of inspection, the presented data are 10 utterances randomly selected from each emotion category except for the neutral category of the AIC test database.}
	\label{fig:emb_tsne}
\end{figure*}

\begin{figure*}[ht]
	\centering
	\begin{minipage}{\linewidth}
		
		\begin{minipage}{0.32\linewidth}
			\centerline{\includegraphics[width=\textwidth]{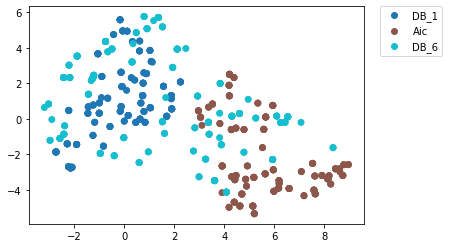}}
			\centerline{(a)}
		\end{minipage}
		\hfill
		\begin{minipage}{0.32\linewidth}
			\centerline{\includegraphics[width=\textwidth]{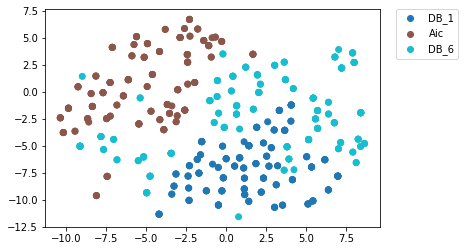}}
			\centerline{(b)}
		\end{minipage}
		\hfill
		\begin{minipage}{0.32\linewidth}
			\centerline{\includegraphics[width=\textwidth]{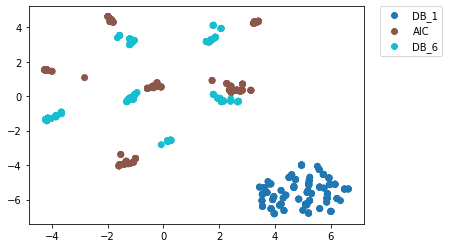}}
			\centerline{(c)}
		\end{minipage}
		\vfill
		%\centering{(a) Surprise}
	\end{minipage}
	\caption{Speaker distribution of the emotion embeddings created by different models (a) GST, (b) VAE, and (c) Proposed EDM module.}
	\label{fig:spk_tsne}
\end{figure*}

In contrast, the proposed method achieves reasonable performance on both identity preservation and emotional expressiveness in terms of all emotion categories. From the point of average performance, the proposed method obtains the best emotion similarity DMOS score, which is 38.4\% and 9.1\% relatively higher than that of Mspk-GST and Mspk-VAE respectively, and comparable speaker similarity DMOS score with that achieved by Mspk-GST, which is only 2.0\% relatively lower than the latter. Furthermore, no significant performance drop appears on either speaker preservation or emotion transfer in terms of any emotion category, indicating that the proposed method can achieve a good balance between maintaining the target speaker's identity and enriching the transferred emotional expression. All these results show the superiority of the proposed method compared to both Mspk-GST and Mspk-VAE, and demonstrate the good performance of the proposed method on the \textit{cross-speaker} emotion transfer TTS task.

\begin{table}[h]
 \caption{Comparison of the proposed method with Mspk-GST and Mspk-VAE in terms of \textbf{source} speaker similarity DMOS with confidence intervals of 95\%. The higher value means more speaker leakage.}
 \label{tab:sourcemos_s}
\setlength{\tabcolsep}{5mm}
 \centering
\begin{tabular}{c|c|c|c}
\toprule
\multicolumn{1}{c|}{Emotion} & \multicolumn{1}{c}{Mspk-GST} & \multicolumn{1}{c}{Mspk-VAE} & \multicolumn{1}{c}{Proposed} \\ \midrule
    fear      &1.32$\pm$0.041  &\bf{1.95}$\pm$\bf{0.053}  &1.42$\pm$0.049  \\
    disgust   &1.28$\pm$0.058  &\bf{2.04}$\pm$\bf{0.048}  &1.38$\pm$0.057  \\
    angry     &1.35$\pm$0.060  &\bf{2.25}$\pm$\bf{0.040}  &1.46$\pm$0.043  \\
    sadness   &1.42$\pm$0.047  &\bf{2.58}$\pm$\bf{0.057}  &1.54$\pm$0.054  \\
    happy     &1.41$\pm$0.055  &\bf{2.62}$\pm$\bf{0.061}  &1.57$\pm$0.049  \\
    surprise  &1.30$\pm$0.052  &\bf{2.27}$\pm$\bf{0.054}  &1.44$\pm$0.061 \\\midrule
    average  &1.35$\pm$0.062  &\bf{2.29}$\pm$\bf{0.058}  &1.47$\pm$0.055 \\
  \bottomrule
\end{tabular}
\end{table}

\begin{table}[h]
 \caption{Cosine similarity of synthesized speech with the target speaker and source speaker.}
 \label{tab:cosine}
\setlength{\tabcolsep}{1.8mm}
 \centering
\begin{tabular}{c|c|c|c|c}
\toprule
\multicolumn{1}{c|}{Speaker} & \multicolumn{1}{c}{Target speaker} & \multicolumn{1}{c}{Mspk-GST} & \multicolumn{1}{c}{Mspk-VAE} & \multicolumn{1}{c}{Proposed} \\ \midrule
    Source speaker &\bf{0.17}   &0.24  &0.36  &0.28  \\\midrule
    Target speaker &\bf{0.75}   &0.65  &0.51  &0.60  \\
  \bottomrule
\end{tabular}
\end{table}

\textbf{The speaker similarity with the source speaker.}
As mentioned in Section~\ref{sc:Introduction}, speaker leakage means that the synthetic speech is mixed with the timbre of the source speaker. To directly show the speaker leakage degree of each model, a DMOS is conducted to evaluate the speaker similarity of the synthetic speech with neutral reference speech of the source speaker, which allows us only to focus on the speaker's timbre rather than emotion categories. In addition to this subjective evaluation, an objective evaluation is conducted to calculate the speaker cosine similarity between synthesized utterances and real neutral utterances from the source speaker and target speaker, respectively. In practice, a pre-trained speaker verification model ECAPA-TDNN~\cite{TDNN} is used to extract the speaker embedding of synthesized speech or real speech. Then cosine similarity is obtained based on the extracted speaker embedding. To avoid the latent effect of semantic information, a pair of speech embeddings are randomly chosen from one model's results and the real speech of the target speaker or source speaker. This randomly sampling method also allows us to obtain the upper bound cosine similarity (0.75) within the target speaker's test set, as well as the lower cosine similarity bound between the target speaker and the source speaker (0.17). These results are presented in Table~\ref{tab:sourcemos_s} and Table~\ref{tab:cosine}, respectively. As can be seen, the results are consistent with the previous conclusion that Mspk-VAE suffers from the worst speaker leakage issue. To be specific, Mspk-VAE's results present the highest similarity to the source speaker, but show the lowest similarity to the target speaker. While the speech synthesized by the proposed method shows a lower similarity with the target speaker than that achieved by Mspk-GST, this gap is not huge, and as explained in the previous, this gap could also be caused by the expressive emotion of speech synthesized by the proposed method.

\subsubsection{Visualization of learned reference embedding}
\label{sc:embedding visualization}

Obtaining speaker-irrelevant emotion is crucial for the \textit{cross-speaker} emotion transfer TTS. A good emotion embedding should be 1) discriminative on distinguishing emotion categories, and 2) indistinguishable for the speaker identities. To compare the ability of emotion embeddings extracted by different models on distinguishing emotion categories or speaker identities, the t-distributed stochastic neighbor embedding (t-SNE)~\cite{Laurens2008Visualizing} method is adopted in this section. The t-SNE visualizes the distribution of embeddings in a two-dimensional space via dimensionality reduction, which allows us to directly present the distribution of embeddings in terms of emotion categories or speaker identities.

\textbf{Distribution in terms of emotion categories}. 
To display the distribution of emotion embeddings extracted by different models, 10 audio utterances of each emotion category from the test set of the source speaker's database are randomly selected, resulting in 60 reference utterances (neutral emotion is not included) which are then embedded as emotion embeddings by different models. The distributions of these embeddings are presented in Fig.~\ref{fig:emb_tsne}. In this figure, each point indicates an emotion embedding, and points with the same color are from the same emotion category. The distance between the two points indicates the relative similarity of the embeddings. Smaller distances indicate more similar embeddings. An ideal emotion encoder should cluster emotion embeddings from the same category close together while embeddings from different categories should be further apart. From this figure, we can see that GST achieves the worst performance on producing the emotion-discriminative embeddings, and many utterances from different emotion categories are mixed together (see Fig.~\ref{fig:emb_tsne} (a)). The distribution turns to better in Fig.~\ref{fig:emb_tsne} (b), which is achieved by VAE. However, the points from the same category are still diverse. In contrast, in Fig.~\ref{fig:emb_tsne} (c), all emotion embeddings from the same emotion category are clustered together, while different clusters are apart from each other, indicating that our proposed EDM module presents the significant superiority in extracting emotion-discriminative embeddings.

\textbf{Distribution in terms of speaker identities}.
Similar to the visualization method in terms of emotion categories, here, we also randomly select 10 utterances from each emotion category except for the neutral emotion, resulting in 60 utterances for the speaker in \textit{AIC}. The same sampling method is used to randomly select 60 utterances from the other emotional database \textit{DB\_6}, and 60 utterances are also randomly selected from \textit{DB\_1} respectively, resulting in 180 reference utterances. Embeddings from the same database are colored by the same color, representing that they are from the same speaker. The visualization results are shown in Fig.~\ref{fig:spk_tsne}. Ideally, good speaker-irrelevant emotion embeddings should contain no speaker identity-related information, which means embeddings of utterances recorded by different speakers should be mixed together in Fig.~\ref{fig:spk_tsne}. As can be seen, the different speaker's embeddings created by GST indeed mix together, while this does not mean that they are good emotion embeddings. The promise of good embeddings is to be emotion-discriminative. Regarding the bad emotion distribution in Fig.~\ref{fig:emb_tsne} (a), the speaker embeddings produced by GST are neither emotion-discriminative nor speaker-discriminative, indicating that these embeddings carry very little information from the reference audio, and that is why Mspk-GST can show that best performance on the target speaker preservation but worst performance on the emotion transfer (see Table \ref{tab:transfer}). For the VAE method, as shown in Fig.~\ref{fig:spk_tsne} (b), there are clear boundaries between clusters that are from different databases (speakers), indicating that the VAE method is unable to produce speaker-irrelevant emotion embeddings. In contrast, for our proposed EDM module (see Fig.~\ref{fig:spk_tsne} (c)) the embeddings from the two emotional datasets (\textit{AIC} and \textit{DB\_6}) are clustered into $6$ clusters and are closed to each other within each cluster, which corresponding to $6$ emotion categories. It is worth noting that because the emotion of speech from \textit{DB\_1} is treated as an independent emotion category, the embeddings from this database are clustered into an independent cluster. This distribution demonstrates that our proposed method can produce speaker-irrelevant emotion embeddings.

\begin{figure}[ht]
	\centering
	\begin{minipage}{\linewidth}
		\begin{minipage}{0.49\linewidth}
			\includegraphics[width=\textwidth]{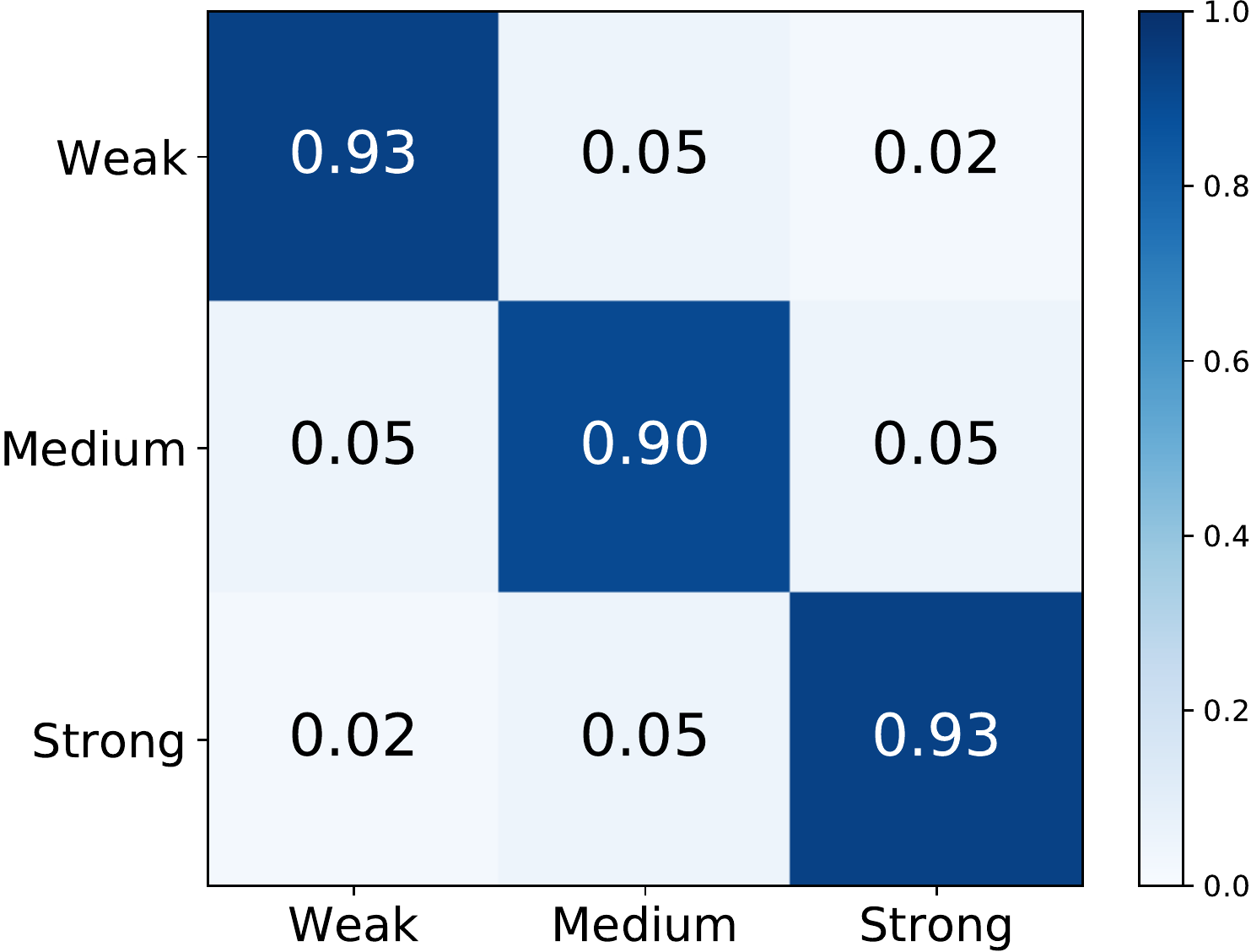}
			\centerline{Fear}
		\end{minipage}
		\vspace{0.1cm}
		\hfill
		\begin{minipage}{0.49\linewidth}
			\includegraphics[width=\textwidth]{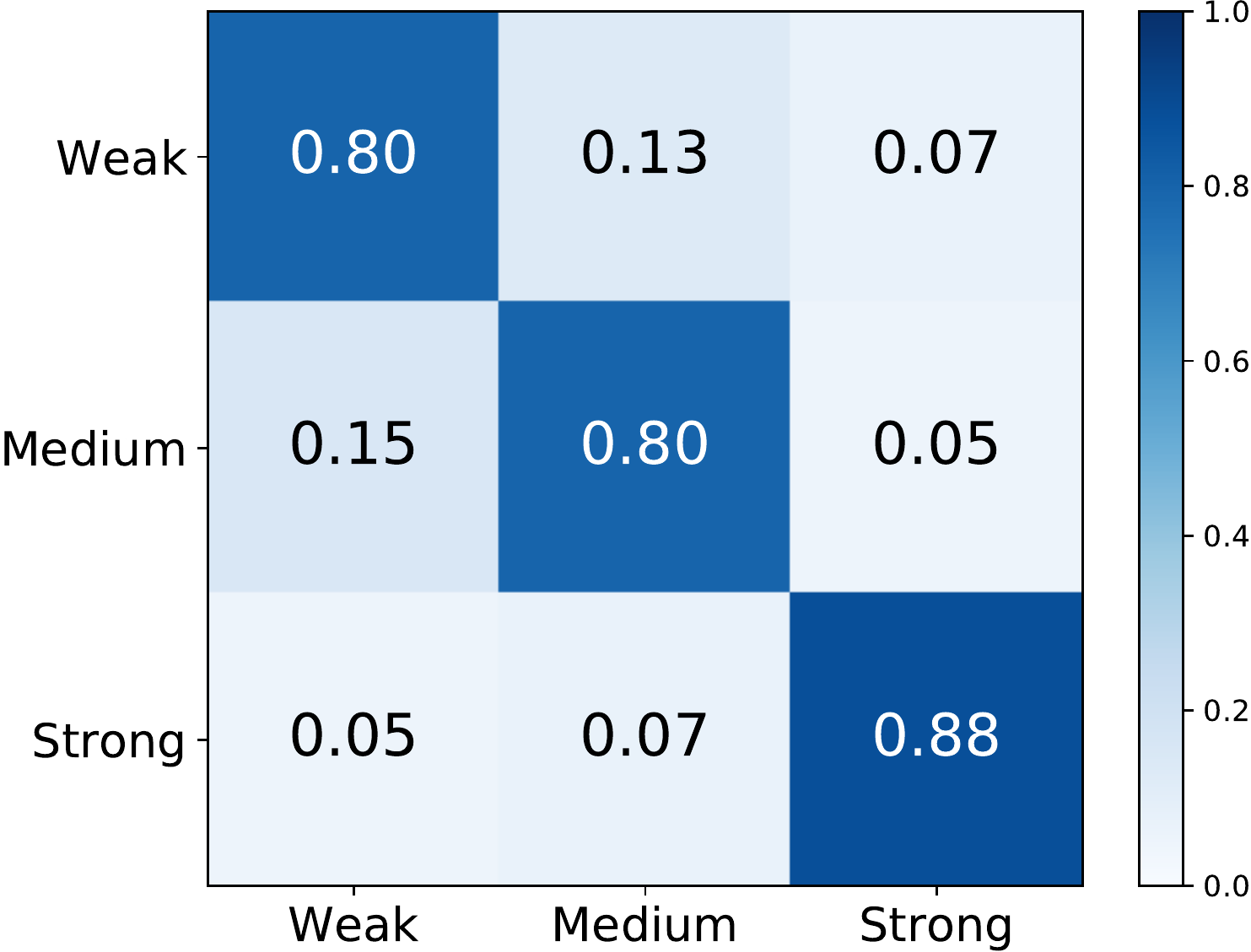}
			\centerline{Disgust}
		\end{minipage}
		\vfill
	\end{minipage}
	
	\begin{minipage}{\linewidth}
		\begin{minipage}{0.49\linewidth}
			\includegraphics[width=\textwidth]{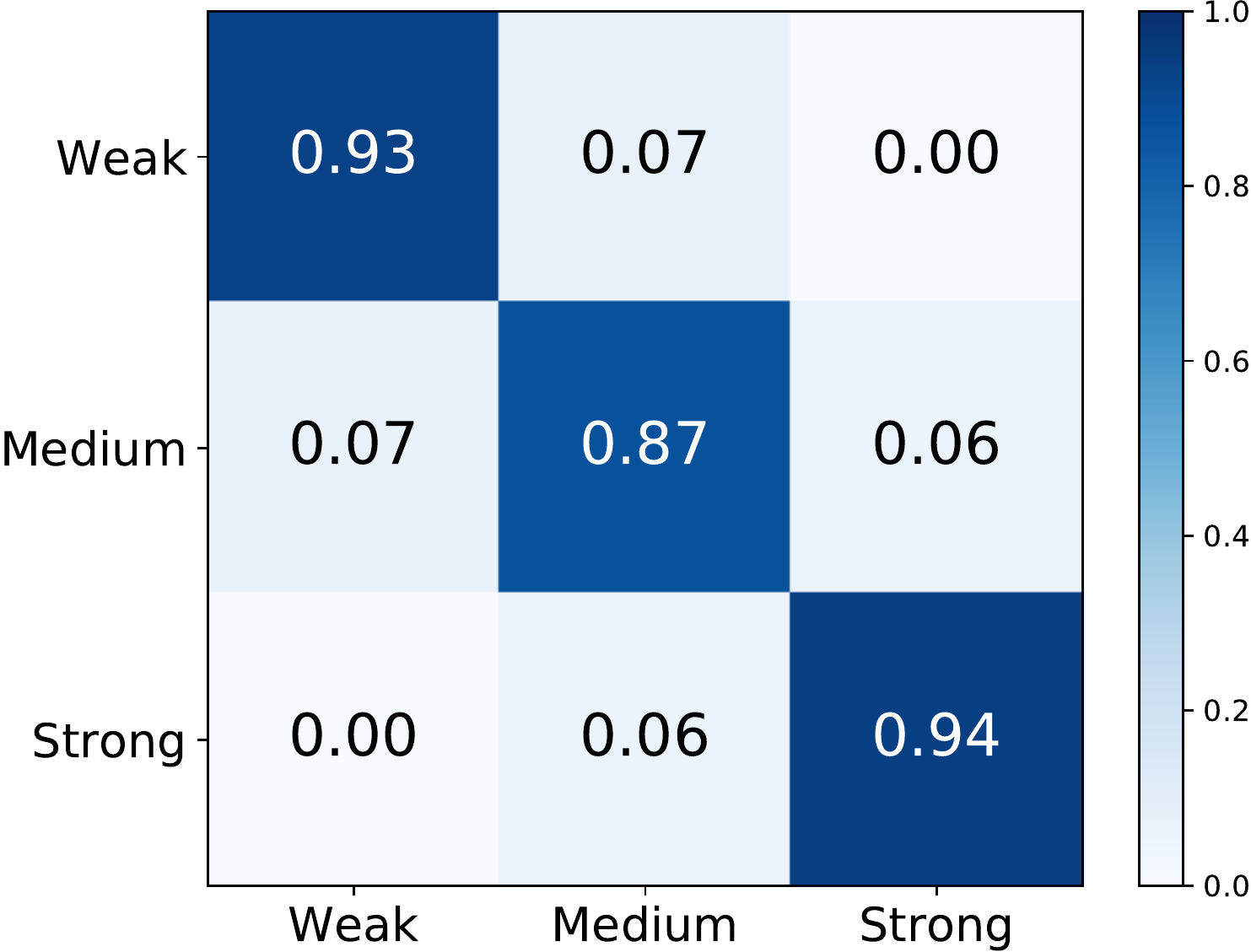}
			\centerline{Angry}
		\end{minipage}
		\vspace{0.1cm}
		\hfill
		\begin{minipage}{0.49\linewidth}
			\includegraphics[width=\textwidth]{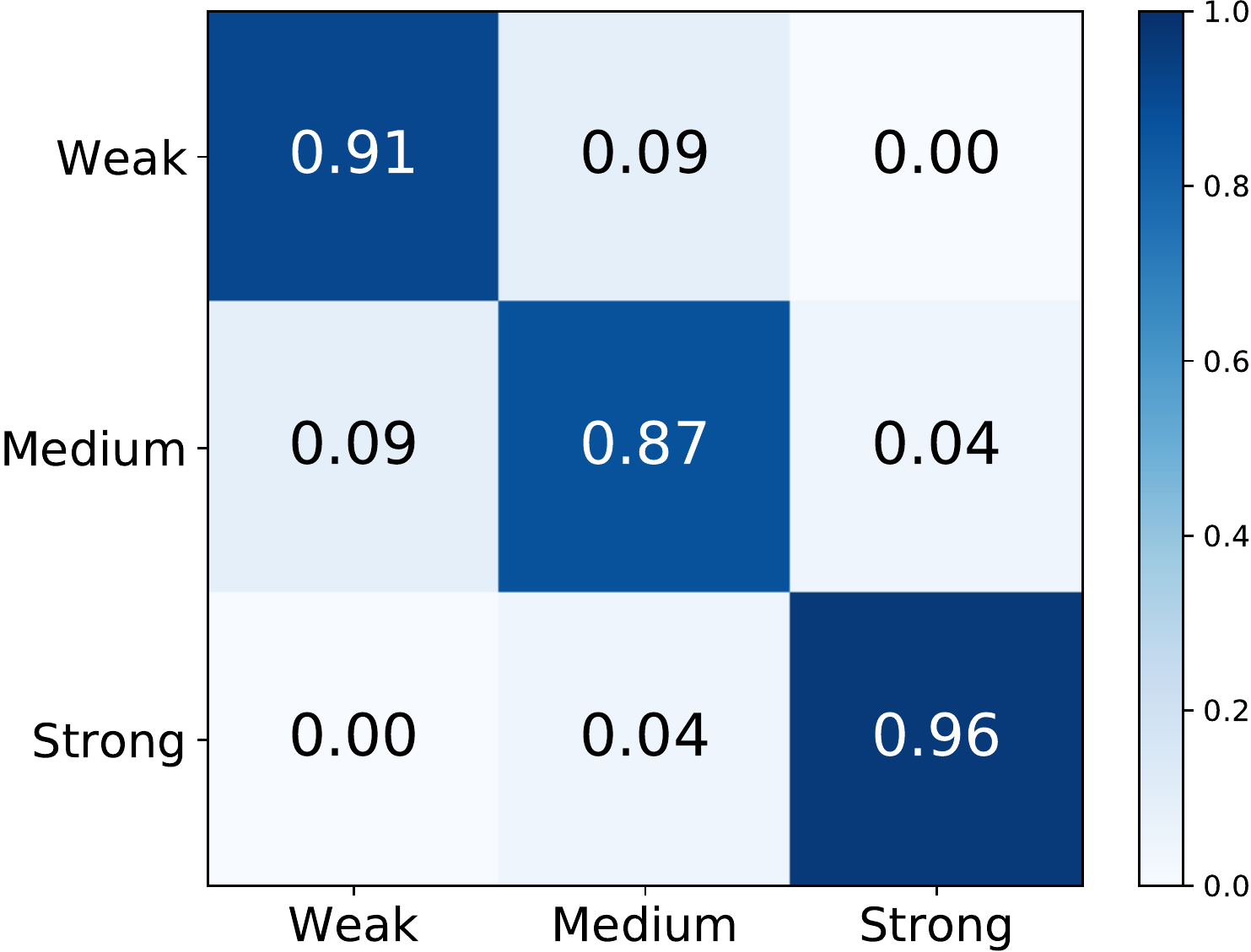}
			\centerline{Sadness}
		\end{minipage}
		\vfill
	\end{minipage}
	\begin{minipage}{\linewidth}
		\begin{minipage}{0.49\linewidth}
			\includegraphics[width=\textwidth]{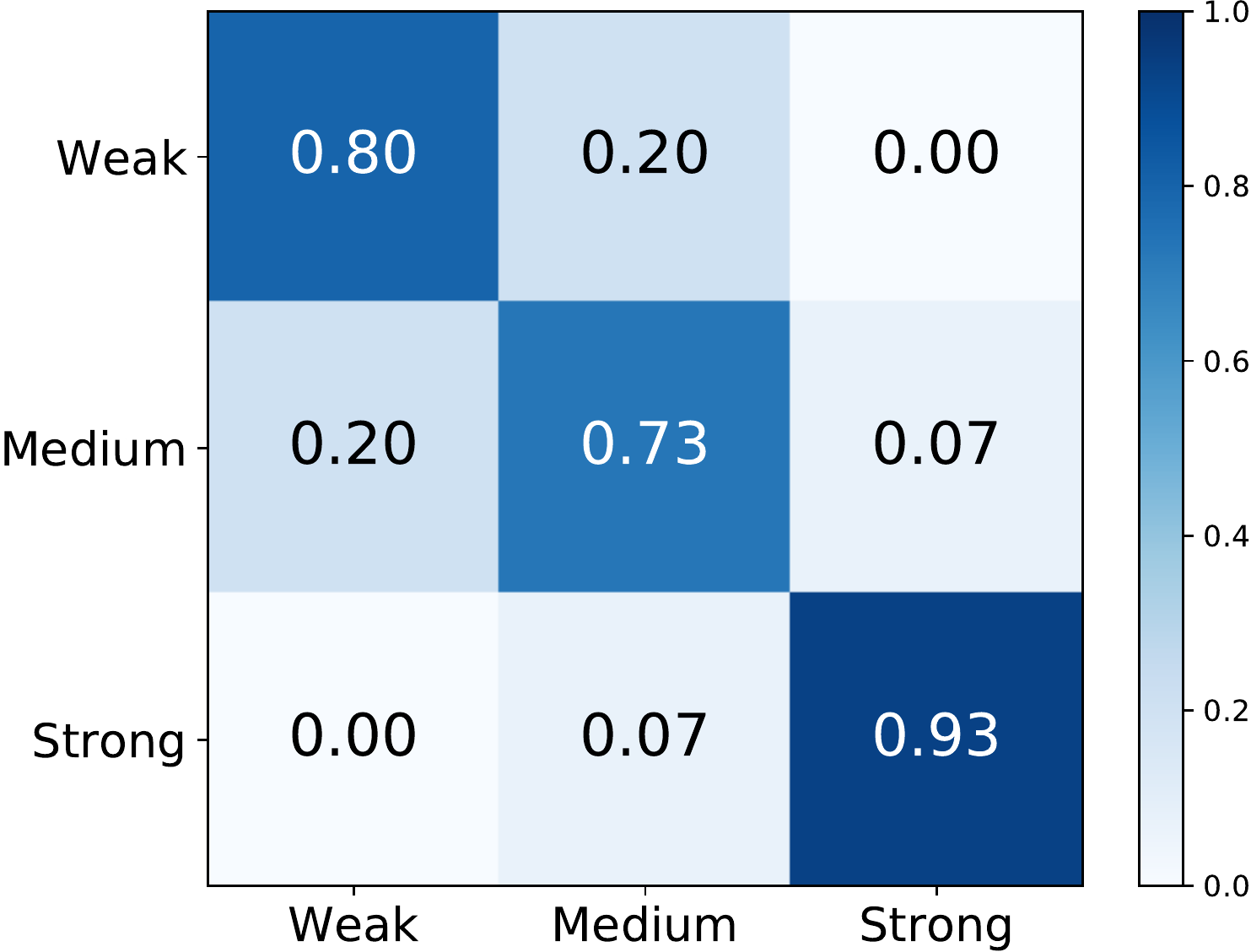}
			\centerline{Happy}
		\end{minipage}
		\hfill
		\begin{minipage}{0.49\linewidth}
			\includegraphics[width=\textwidth]{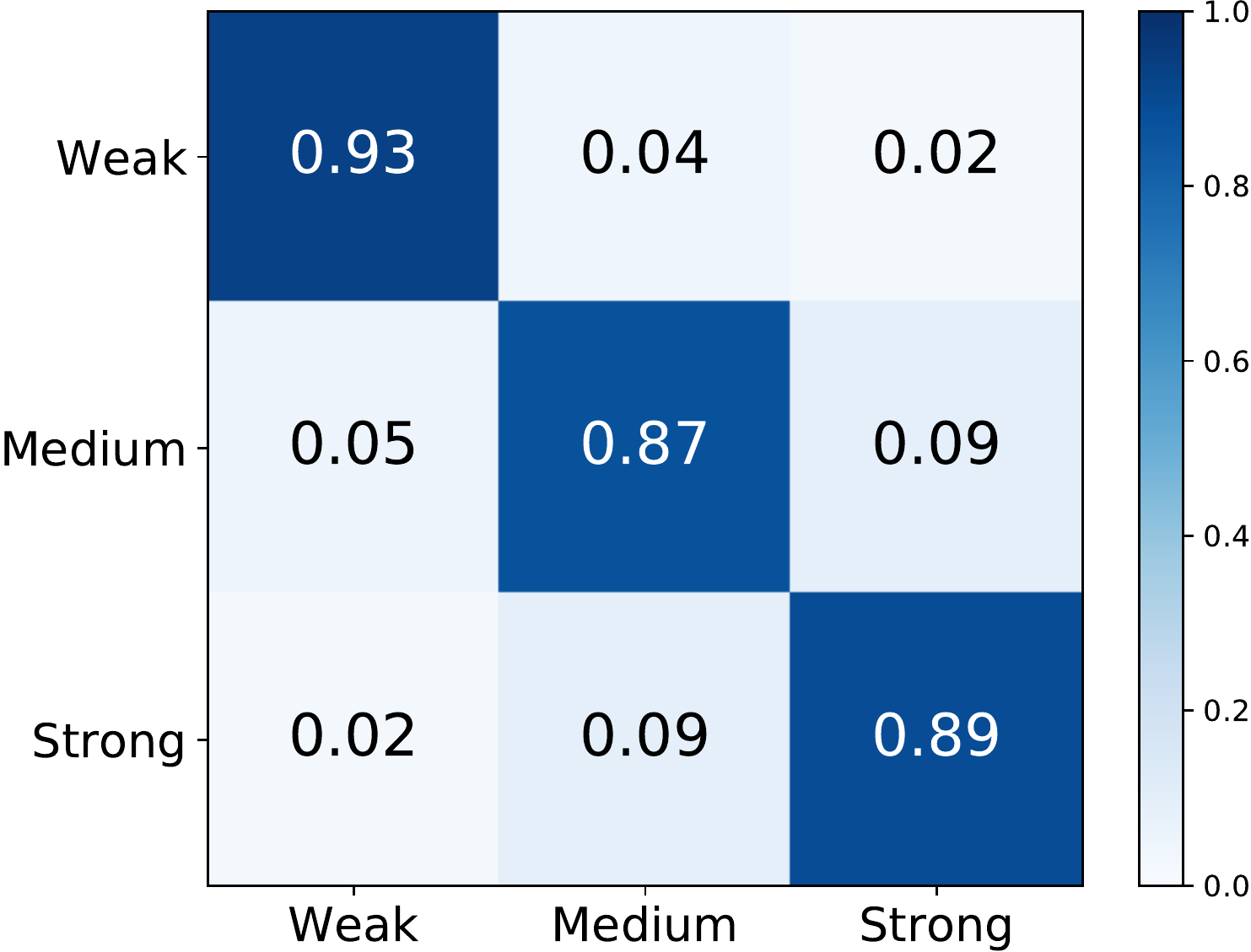}
			\centerline{Surprise}
		\end{minipage}
		\vfill
		%		\centering{(b) DIG-Tacotron2}
	\end{minipage}
	
	\caption{Confusion matrices of synthesized speech from the proposed model. The X-axis and Y-axis of subfigures represent perceived and ground-truth emotion strength, respectively.}
	\label{fig:confuse}
\vspace{-0.5cm} 
\end{figure}

All above experiments and results, including the DMOS scores of the TTS results and the embedding distribution visualization, indicate that the proposed method is superior on the \textit{cross-speaker} emotion transfer TTS task compared to the GST and VAT-based methods.

\begin{figure*}[h]
	\centering
	\centerline{\includegraphics[width=0.56\textwidth]{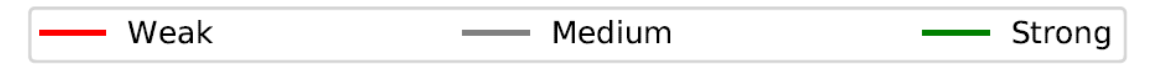}}
    \hfill
	\begin{minipage}{\linewidth}
		
		\begin{minipage}{0.32\linewidth}
			\centerline{\includegraphics[width=\textwidth]{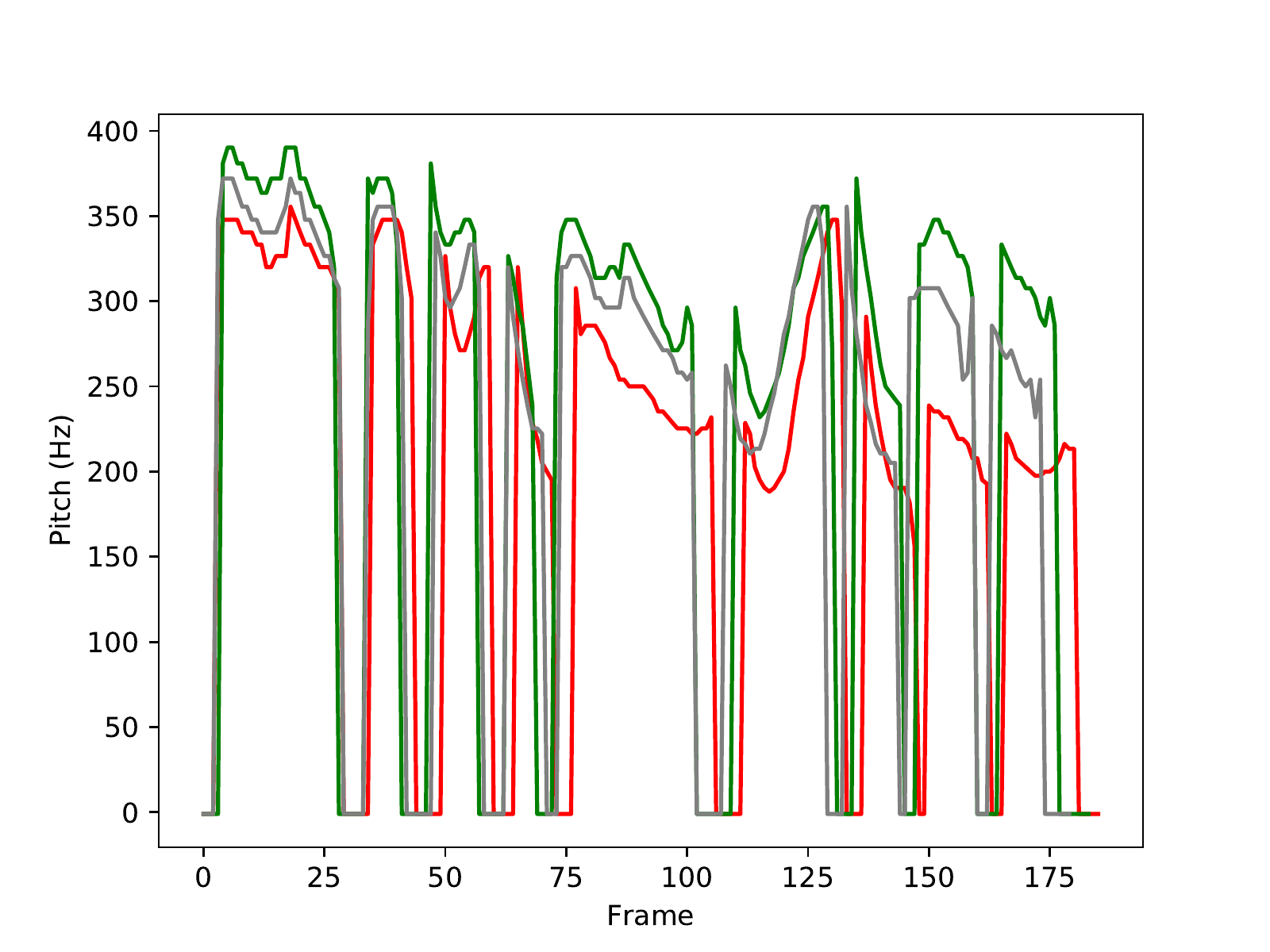}}
			\centerline{Fear}
		\end{minipage}
		\hfill
		\begin{minipage}{0.32\linewidth}
			\centerline{\includegraphics[width=\textwidth]{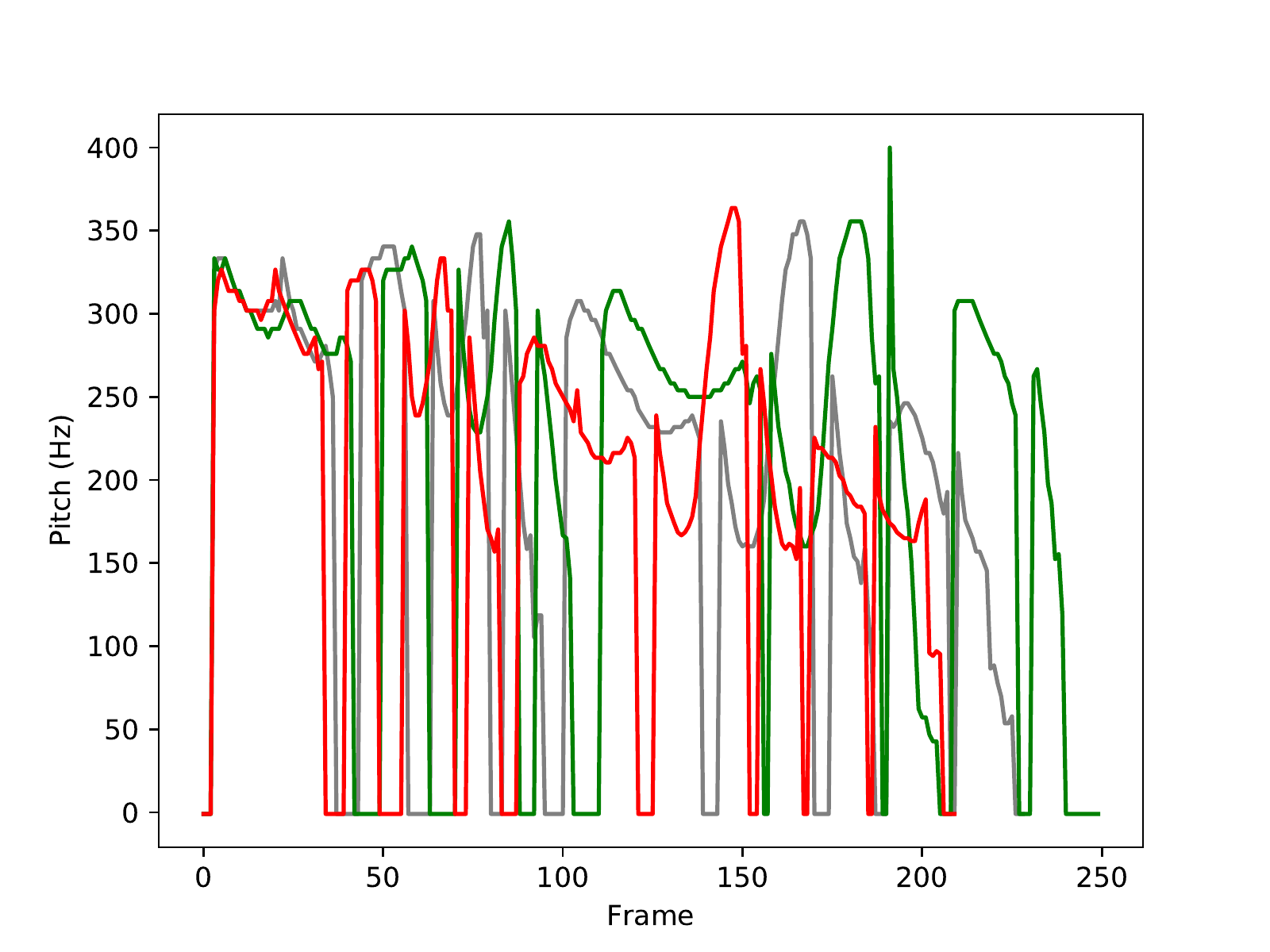}}
			\centerline{Disgust}
		\end{minipage}
		\hfill
		\begin{minipage}{0.32\linewidth}
			\centerline{\includegraphics[width=\textwidth]{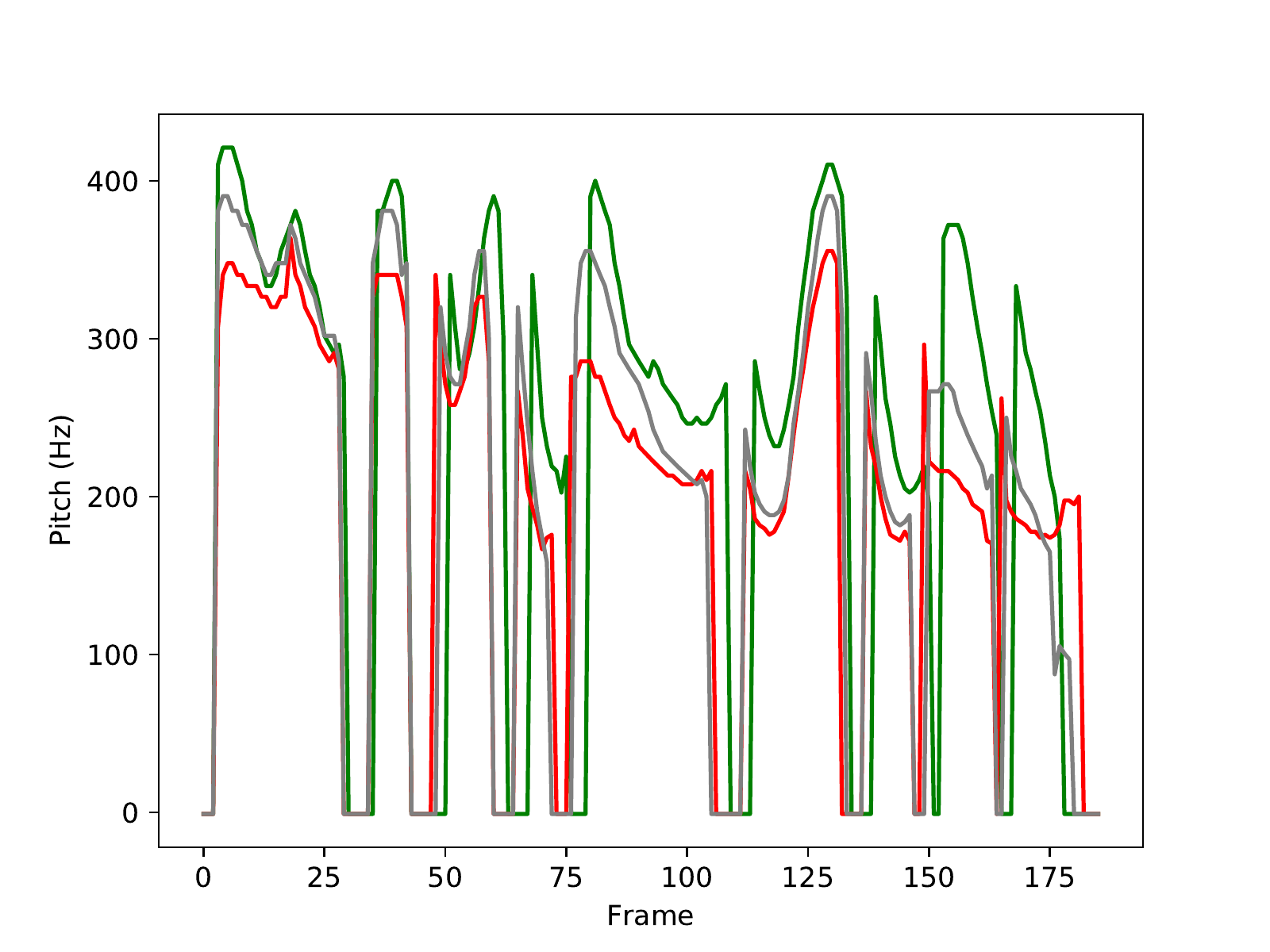}}
			\centerline{Angry}
		\end{minipage}
		\vfill
		%\centering{(a) Surprise}
	\end{minipage}
	\begin{minipage}{\linewidth}
		
		\begin{minipage}{0.32\linewidth}
			\centerline{\includegraphics[width=\textwidth]{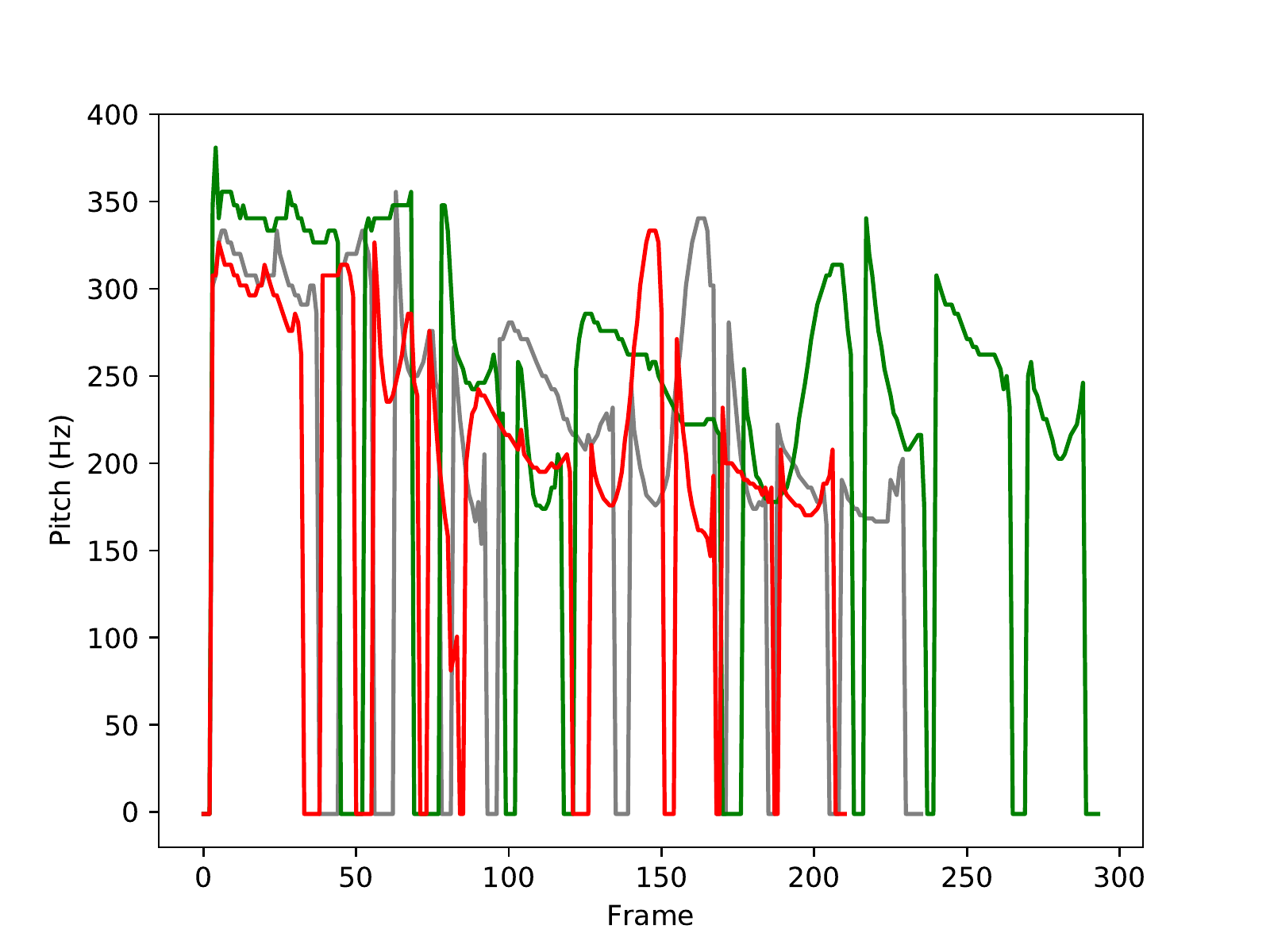}}
			\centerline{Sadness}
		\end{minipage}
		\hfill
		\begin{minipage}{0.32\linewidth}
			\centerline{\includegraphics[width=\textwidth]{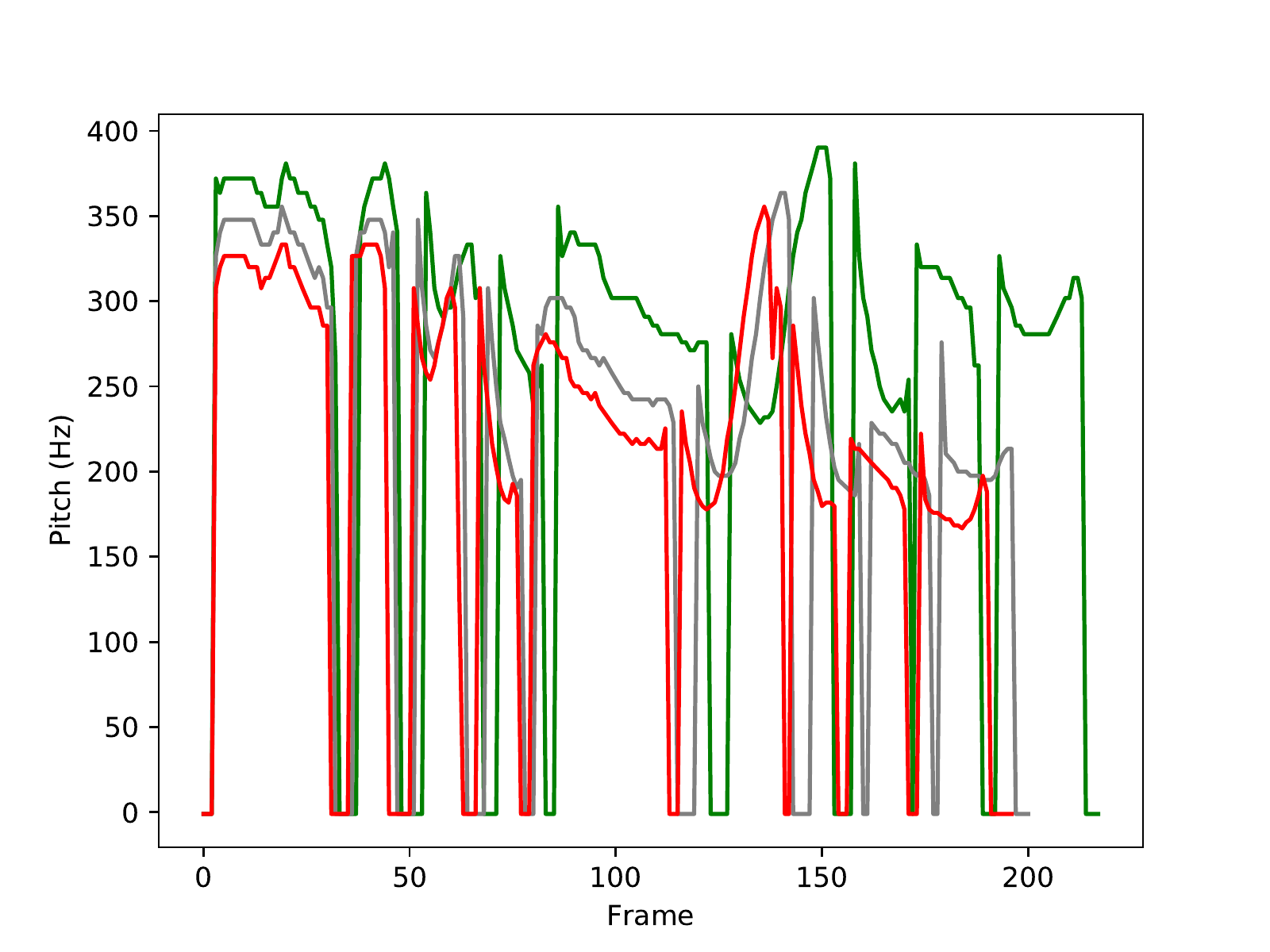}}
			\centerline{Happy}
		\end{minipage}
		\hfill
		\begin{minipage}{0.32\linewidth}
			\centerline{\includegraphics[width=\textwidth]{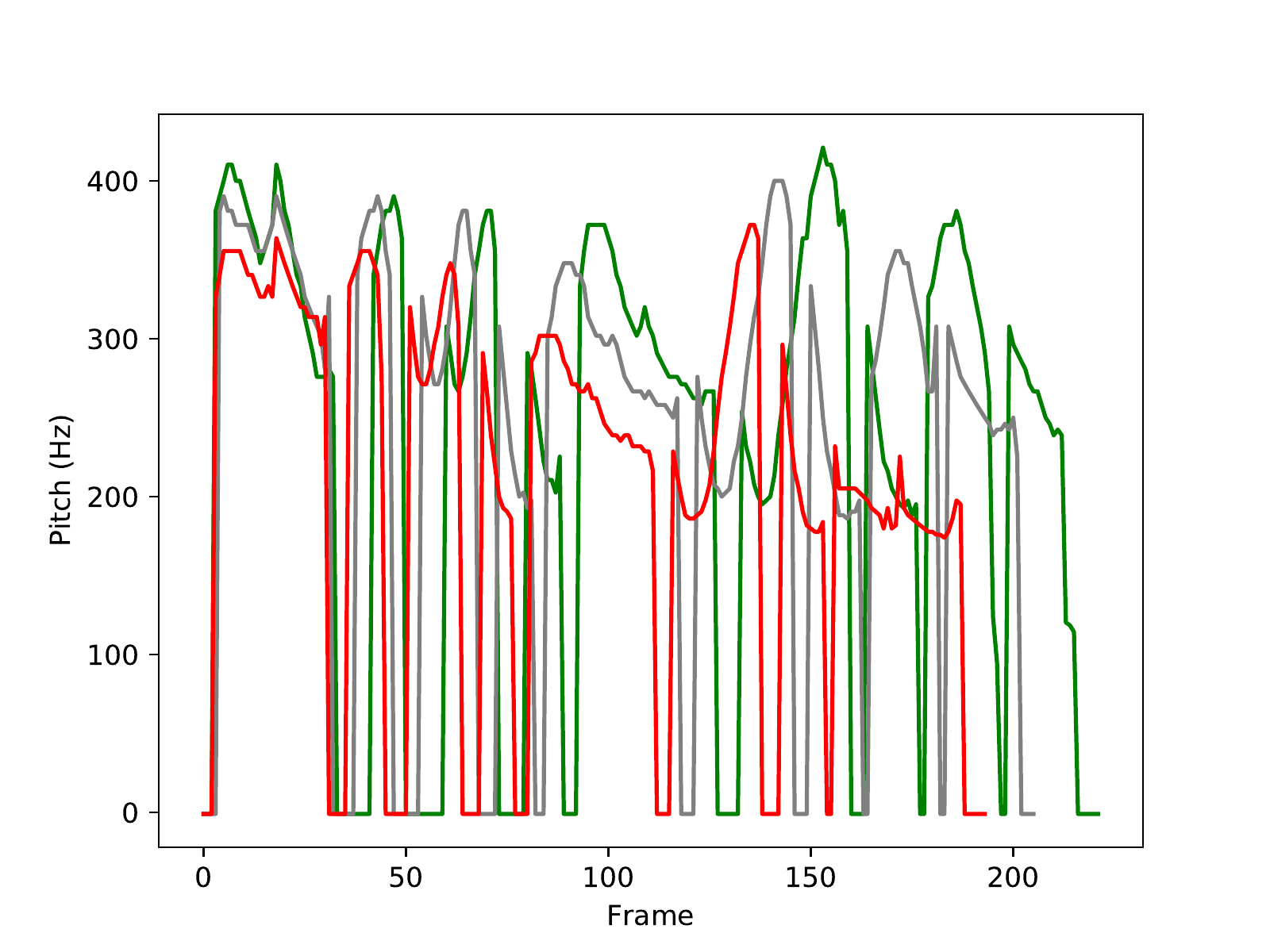}}
			\centerline{Surprise}
		\end{minipage}
		\vfill
		%\centering{(a) Surprise}
	\end{minipage}
	\caption{Pitch contours of synthesized speech with the same text for six emotions and three emotion strengths.}
	\label{fig:pitch}
\end{figure*}

\subsection{Cross-speaker emotion strength control}
\label{sc:emtion_control}
In this section, we show the ability of the proposed method on controlling the transferred emotion strength. To this end, speech with three transferred emotion strength levels, i.e., weak, medium, and strong, is synthesized for each input sentence, 180 utterances will be synthesized. To evaluate the relative strength of synthetic speech synthesized with different emotion controller scalars, a human perceptual ranking experiment was conducted. Specifically, for each sentence, given three synthetic speech with different emotion strengths but the same emotion category, participants are asked to sort them according to the emotion strength, i.e., ranking speech to the strength order of weak, medium, and strong. In this experiment, 20 participants participated, and each participant was shown with 60 groups of synthetic speech. Note that although the participants were asked to rank three synthetic speech utterances in each group, we treat the evaluation as a classification task. To be specific, if synthesized speech controlled by a weak scalar is ranked in the first place (with the label of weak), we treat it as a correctly classified sample (correctly reflect the emotion strength) regardless of the order of the other two utterances in the same group.

The ranking results are drawn in a confusion matrix, as shown in Fig~\ref{fig:confuse}, for each emotion category. In this figure, a value on the diagonal means the accuracy of that synthetic sample correctly reflects the emotion strength as expected. As we can see, for all emotion categories, the synthetic utterances achieve high accuracies in terms of all strength levels. For instance, for the emotion of \textit{fear}, all the classification accuracies are no less than 90\%. Even for the case with the lowest accuracy, i.e., Medium for the \textit{happy}, the accuracy is still larger than 70\%, indicating that the proposed method can successfully control the emotion strength performed by the target speaker.

\begin{table}[h]
 \caption{Speaker similarity DMOS for synthetic speech with three typical emotion strengths, i.e.,  weak, medium, and strong, with confidence intervals of 95$\%$. The higher value means better performance, and the bold indicates the best performance out of three emotion strengths in terms of each emotion.}
 \label{tab:controlmos}
\setlength{\tabcolsep}{5mm}
 \centering
\begin{tabular}{l|l|l|l}
\toprule
\multicolumn{1}{c|}{Strength} & \multicolumn{1}{c}{Weak} & \multicolumn{1}{c}{Medium} & \multicolumn{1}{c}{Strong} \\ \midrule
    fear      &\bf{3.89}$\pm$\bf{0.073}  &3.79$\pm$0.071  &3.68$\pm$0.068  \\
    disgust   &\bf{4.01}$\pm$\bf{0.075}  &3.87$\pm$0.078  &3.84$\pm$0.072  \\
    angry     &\bf{3.90}$\pm$\bf{0.072}  &3.82$\pm$0.07  &3.63$\pm$0.063  \\
    sadness   &\bf{3.79}$\pm$\bf{0.081}  &3.72$\pm$0.072  &3.59$\pm$0.082  \\
    happy     &\bf{3.76}$\pm$\bf{0.073}  &3.79$\pm$0.076  &3.70$\pm$0.069  \\
    surprise   &\bf{3.90}$\pm$\bf{0.079}  &3.75$\pm$0.076  &3.67$\pm$0.074 \\\midrule
    average  &\bf{3.89}$\pm$\bf{0.075}  &3.79$\pm$0.073  &3.68$\pm$0.072 \\
  \bottomrule
\end{tabular}
\vspace{-0.1cm}
\end{table}

\textbf{Prosody Diversity}. The emotion strength of speech has a close relation to prosody. Here, in addition to the subjective evaluation, the prosody aspects of synthetic speech with different transferred emotion strengths are also compared. The visualized comparisons are shown in Fig~\ref{fig:pitch}, in which the pitch trajectory of each synthetic speech is drawn based on the Mel-spectrograms. All the presented synthetic speech utterances are based on the same input sentence. For each emotion category, the presented synthetic speech utterances are with different emotion strengths, i.e., weak, medium, and strong. As can be seen, in each subfigure, the pitch trajectories of different strengths present a similar trend but with different peaks and duration. For instance, the pitch and tone variation of $surprise$ and $anger$ increase with the increase of strength from weak to strong. As for $sadness$ and $disgust$, the speaking rate slows down as the strength increases, accompanied by narrow pitch variation, indicating the significant effect of our proposed method on adjusting the emotion strength, and also the ability of the proposed method on synthesizing diversity prosodies.

\begin{table*}[]
 \caption{DMOS scores for speaker and emotion similarity of ablation studies with confidence intervals of 95$\%$.}
 \label{tab:ablation}
\setlength{\tabcolsep}{6mm}
 \centering
\begin{tabular}{l|lll|lll}
\toprule
\multicolumn{1}{c|}{\multirow{2}{*}{Emotion}} & \multicolumn{3}{c|}{Speaker similairty DMOS}                                                  & \multicolumn{3}{c}{Emotion similarity DMOS}                                                \\ \cmidrule{2-7} 
\multicolumn{1}{c|}{}                         & \multicolumn{1}{c}{w/o 2ort} & \multicolumn{1}{c}{w/o ort} & \multicolumn{1}{c|}{Proposed} & \multicolumn{1}{c}{w/o 2ort} & \multicolumn{1}{c}{w/o ort} & \multicolumn{1}{c}{Proposed} \\ \midrule
 fear        &1.60$\pm$0.069  &3.13$\pm$0.06   &\bf{3.95}$\pm$\bf{0.062}  &\bf{4.05}$\pm$\bf{0.082}  &4.01$\pm$0.078  &3.75$\pm$0.073 \\
 disgust     &1.22$\pm$0.04   &3.45$\pm$0.049  &\bf{4.01}$\pm$\bf{0.055}  &\bf{3.69}$\pm$\bf{0.076}  &3.65$\pm$0.07   &3.55$\pm$0.075 \\
 angry       &1.58$\pm$0.076  &3.40$\pm$0.058  &\bf{3.90}$\pm$\bf{0.056}  &\bf{4.05}$\pm$\bf{0.074}  &3.77$\pm$0.065  &3.62$\pm$0.07 \\
 sadness     &1.61$\pm$0.085  &2.97$\pm$0.077  &\bf{3.79}$\pm$\bf{0.064}  &\bf{4.35}$\pm$\bf{0.081}  &4.14$\pm$0.076  &3.94$\pm$0.055  \\
 happy       &1.40$\pm$0.052  &2.90$\pm$0.042  &\bf{3.76}$\pm$\bf{0.063}  &\bf{3.88}$\pm$\bf{0.077}  &3.73$\pm$0.063  &3.70$\pm$0.06   \\
 surprise    &1.65$\pm$0.06   &3.45$\pm$0.058  &\bf{3.90}$\pm$\bf{0.055}  &\bf{3.83}$\pm$\bf{0.086}  &3.75$\pm$0.069  &3.69$\pm$0.068   \\\hline
  \textbf{average}  &1.51$\pm$0.062 &3.22$\pm$0.06 &\bf{3.89}$\pm$\bf{0.057}  &\bf{3.98}$\pm$\bf{0.079} & 3.84$\pm$0.07 &3.71$\pm$0.066 \\
  \bottomrule
\end{tabular}
\end{table*}

\begin{table}[h]
 \caption{Emotion strength control CMOS in the \textit{same-speaker} scenario for comparison the emotional expressiveness of three strengths synthesized speech with the corresponding emotional reference audio.}
 \label{tab:cmos}
\setlength{\tabcolsep}{3.5mm}
 \centering
\begin{tabular}{l|c|l|c|l}
\toprule
\multicolumn{1}{c|}{Emotion} & \multicolumn{1}{c}{Rererence} & \multicolumn{1}{c}{Weak} & \multicolumn{1}{c}{Medium}  & \multicolumn{1}{c}{Strong} \\ \midrule
    fear     &0 &-0.58 &0.61 &1.72  \\
    disgust  &0 &-0.64 &0.83 &1.57   \\
    angry    &0 &-0.61 &0.55 &1.66  \\
    sadness  &0 &-0.45 &0.65 &1.75  \\
    happy    &0 &-0.47 &0.58 &1.67  \\
    surprise  &0 &-0.88 &0.56 &1.62 \\\midrule
    average   &/ &-0.61 &0.63 &1.67 \\
  \bottomrule
\end{tabular}
\end{table}

\textbf{Effect on speaker identity preservation}.
The proposed method for the controllable emotion transfer TTS raises a question: whether controlling the emotion strength with scalar would affect the identity preservation of the target speaker? To answer this question, the speaker similarity DMOS test was performed for synthetic speech with different emotion strengths. The results are shown in Table~\ref{tab:controlmos}. As can be seen, it indeed shows a trend that stronger emotion transfer tends to bring lower speaker similarity DMOS. However, this DMOS difference between speech with different transferred emotion strengths is not significant. For instance, for the DMOS value averaging overall emotion categories, the score achieved with the strong emotion is only 5.4\% relatively lower than that with weak emotion, indicating that the effect of the emotion control on the speaker preservation is acceptable.

\textbf{Emotion strength control in the \textit{same-speaker} scenario.}

While the previous experiments have shown that the proposed method indeed can control the emotion strength, the lack of comparison with real reference speech makes it hard to tell whether the proposed method can produce emotional speech that is stronger or weaker than the reference speech. To further evaluate the control ability of the proposed method, a comparison means opinion score (CMOS) test was performed by comparing the reference speech and synthetic speech that synthesized with different strength scales. In practice, the target speaker is no longer the speaker without emotional recordings but the same as the source speaker that provides reference speech, which allows us only to focus on the emotion strength rather than emotion categories or speaker information. The CMOS ranges from -3 to 3 in 0.5 point increments, and score 0 means the synthetic utterance and the reference have the same emotion strength. A positive value means the synthetic utterance has a stronger emotion than the reference and vice versa.

Table~\ref{tab:cmos} reports CMOS results, where the score of the reference is fixed to 0. As can be seen, the synthetic speech can be obviously controlled with emotion strength that is weaker or stronger than the reference speech, indicating the effective ability of the proposed method on emotion strength control.

\subsection{Ablation studies}
Compared to our preliminary work \cite{Li2021ControllableET}, the orthogonal constraint between the emotion embedding and speaker embedding is the main extended method in this paper. To evaluate the effectiveness of this proposed method, an ablation study is performed. Specifically, two variants are evaluated: 1) no orthogonal constraint is adopted neither for the emotion embedding constraint for the Tacotron's encoder input, nor for the speaker embedding constraint for the Tacotron's decoder output. We denote this variant as ``w/o 2ort''; 2) the orthogonal constraint is only adopted for the emotion embedding for the Tacotron's encoder input, but not for the speaker embedding constraint for the Tacotron's decoder output. We denote this variant as ``w/o ort''.

Performance comparison of these two variants and the proposed method is shown in Table~\ref{tab:ablation}. For speaker similarity, the variant of ``w/o 2ort'' gets the lowest speaker similarity for all evaluated emotion types, which could be caused by the reason that the extracted emotion embedding contains source speaker-related information. By removing the speaker information with orthogonal constraint, ``w/o ort'' outperforms ``w/o 2ort'' for all the emotion types. However, due to that, there is no explicit constraint on the decoder to ensure sufficient speaker and emotion discrimination, making the speaker similarity DMOS still has a large gap to the performance of our proposed method.

For the emotion similarity, while the emotional expression of ``w/o 2ort'' and ``w/o ort'' is better than the proposed method, these performance differences are not significant. Actually, in the empirical observation, we found that the proposed method achieves reasonable performance in terms of speaker similarity and emotion expression. The DMOS ratings in most emotion categories are close to $good$, except for the $disgust$ category, for which the expression is extremely related to the timbre of the source speaker, and the expression of $disgust$ in the training data is similar to neutral. These results demonstrate that the proposed orthogonal constraint method with the help of the speaker encoder learning method and emotion encoder learning method is effective, and the proposed emotion transfer TTS model is well designed.

\section{Discussion}
\label{sc:discussion}

Automatically synthesizing emotional expressive speech has great potential in many human-computer interaction scenarios. While several efforts have been performed towards the emotion transfer TTS, no previous work exists for the \textit{cross-speaker} emotion transfer and strength control TTS. In this paper, for the first time, a \textit{cross-speaker} emotion transfer and strength control TTS method is proposed. Extensive experiments show that this proposed method is able to synthesize reasonable speech with the emotion transferred from another source speaker, and the emotion strength can be controlled reasonably and flexibly. 

The emotion disentangling module (EDM) is an important module proposed in this paper to obtain the speaker-irrelevant emotion embedding. The speaker encoder in EDM trained with the speaker classification loss and the reversal emotion classification loss provides the emotion-irrelevant speaker embedding that is used to constrain the emotion encoder via the orthogonal loss. The ablation study demonstrates the importance of this orthogonal constraint and indicates the good design of the EDM. 

Using reference audios with different emotion strengths is a straightforward way to perform emotion strength control. However, as mentioned in Section~\ref{sc:Introduction}, it is not trivial to build such a database with not only emotion categories but also emotion strengths. Furthermore, it is not easy to manually select a proper reference to deliver the expected emotion strength even if such a database exists. In this work, the control of the transferred emotion strength is realized by a flexible scalar value that is multiplied by the emotion embedding, making it can bypass the dependence on the emotion strength of the reference audio. The experimental results show that this method can successfully adjust the emotion strength of synthesized speech, and meanwhile brings little impact on the speaker preservation. While the reasonable controllable performance has been proved by the experiments, an objective explanation behind this phenomenon, especially why the scalar has a similar effect to different emotion categories should be further explored in the future.

In the current work, only one speaker is adopted as the target speaker, with the speaker controller and EDM, it is easy to extend our method to multi-target speakers when multi-speakers databases are available during the training processing. However, when a speaker's database is inaccessible during the training stage, which is called unseen target speaker, this speaker can not be a target speaker in our system. It could be an interesting topic for future research to develop a TTS system with the ability to transfer the emotion from the source speaker to the unseen speaker who doesn't exist in the training data. Furthermore, the emotion strength in the current work is controlled at the global level. However, the emotion expressions of human speech are hierarchical in nature. For instance, some keywords are expected to strongly convey the emotion than the rest. Therefore, how to control speech emotion hierarchically could be an interesting topic in the future.

\section{Conclusion}
\label{sc:conclusion}

This paper proposes a controllable cross-speaker emotion transfer method based on Tacotron2. In order to solve the speaker leakage problem, an emotion disentangling module (EDM) is designed to obtain the speaker-irrelevant emotion embeddings. It consists of an emotion encoder and a speaker encoder. The speaker encoder is trained to obtain emotion-irrelevant embeddings, with which we can constrain the emotion embedding to be speaker-irrelevant via an orthogonal loss. Moreover, to deliver the emotion and meanwhile preserve the target speaker identity, the EDM is also used to explicitly ensure the speaker identity and emotion category in the synthesized Mel-spectrograms to be that as expected. The control of the emotion strength is realized by a flexible scalar value. Experimental results demonstrate that the proposed method achieves good performance in transferring emotions and meanwhile preserving the voice of the target speaker, and it can produce synthetic speech with diverse prosody for the target speaker with strength control.

\bibliographystyle{IEEEtran}
\bibliography{mybibfile.bib}

% \end{IEEEbiography}



\end{document}